\documentclass[useAMS,usenatbib,fleqn,usedcolumn]{mnras}

\usepackage[british]{babel}             
\usepackage{graphicx}                   

\usepackage{bm}
\usepackage{amsmath}
\usepackage{amssymb}

\newcommand{\unit}[1]{\ensuremath{\,\mathrm{#1}}}
\newcommand{\diff}{\ensuremath{\mathrm{d}}}
\newcommand{\deriv}[2]{\ensuremath{\frac{\diff #1}{\diff #2}}}
\newcommand{\tderiv}[2]{\ensuremath{\frac{\diff #1}{\diff #2}}}
\newcommand{\St}{\ensuremath{\mathrm{St}}}
\newcommand{\reply}[1]{#1}

\title[Breakthrough]{Breakthrough revisited: investigating the requirements for growth of dust beyond the bouncing barrier.}
\author[Booth et al.]{Richard A. Booth\thanks{E-mail: rab200@ast.cam.ac.uk}$^1$, Farzana Meru$^1$, Man Hoi Lee$^{2}$, Cathie J. Clarke$^{1}$ \\
$^{1}$Institute of Astronomy, University of Cambridge, Madingley Road, Cambridge, CB3 0HA, United Kingdom \\
$^{2}$Department of Earth Sciences and Department of Physics, The University of Hong Kong, Pokfulam Road, Hong Kong}

\begin{document}

\date{Accepted . Received ; in original form }

\pagerange{\pageref{firstpage}--\pageref{lastpage}} \pubyear{2017}

\maketitle

\label{firstpage}

\begin{abstract}
For grain growth to proceed effectively and lead to planet formation a number of barriers to growth must be overcome. One such barrier, relevant for compact grains in the inner regions of the disc, is the `bouncing barrier' in which large grains ($\sim\unit{mm}$ size) tend to bounce off each other rather than sticking. However, by maintaining a population of small grains it has been suggested that cm-size particles may grow rapidly by sweeping up these small grains. We present the first numerically resolved investigation into the conditions under which grains may be lucky enough to grow beyond the bouncing barrier by a series of rare collisions leading to growth (so-called `breakthrough'). \reply{Our models support previous results, and show that in simple models breakthrough requires the mass ratio at which high velocity collisions transition to growth instead of causing fragmentation to be low,  $\phi \lesssim 50$. However, in models that take into account the dependence of the fragmentation threshold on mass-ratio, we find breakthrough occurs more readily, even if mass transfer is relatively inefficient. This suggests that bouncing may only slow down growth, rather than preventing growth beyond a threshold barrier. However, even when growth beyond the bouncing barrier is possible, radial drift will usually prevent growth to arbitrarily large sizes.} 
\end{abstract}


\section{Introduction}
\label{Sec:Intro}

Dust evolution is an essential component of disc evolution and planet formation. For grains smaller than about a millimetre, growth is believed to be efficient as collisions occur at low velocity and typically result in sticking, with growth rates slowing at larger sizes \citep{Dullemond2005}. However, this predicts a rapid removal of small grains, which is in conflict with observations that show they remain abundant throughout the disc's lifetime \citep{Haisch2001,Cieza2007,Su2009}. Furthermore, we know that discs simultaneously harbour grain sizes up to at least cm sizes throughout their lifetimes \citep{Wilner2005,Rodmann2006,Ricci2010,Miotello2014}. 

However, as grains grow they begin to decouple from the background flow and their relative velocities increase, initially through random velocities driven by turbulence \citep{Volk1980,Markiewicz1991,Ormel2007} and later as they drift towards the star \citep{Weidenschilling1977}.  At high enough velocities collisions between grains tend to result in fragmentation ($\gtrsim 1\unit{m\,s}^{-1}$ for silicates, see \citealt{Blum2008}). While fragmentation provides a mechanism by which the population of small grains can be replenished, it also slows down or prevents growth beyond a certain size. \reply{Although fragmentation reduces the loss of grains via migration, by limiting growth it does not resolve the problem of forming large particles by coagulation}. However, the radial drift barrier may be overcome at pressure maxima which can trap dust grains, such as at gaps induced by massive planets  \citep{Fouchet2007,Ayliffe2012,Zhu2012,Lambrechts2014,Rosotti2016}, in vortices \citep{Barge1995,Adams1995,Tanga1996,Johansen2004,Lyra2009}, or \reply{perhaps at evaporation fronts \citep{Estrada2016,Stammler2017,Schoonenberg2017}} it is not clear how such planet induced traps can explain the first planets to form.

With the recent resurgence of theories in which the dust's own self-gravity drives growth directly to the required km-sizes, perhaps enhanced via drag driven instabilities such as the streaming instability \citep{Youdin2005,Johansen2007,Bai2010,Simon2016} or taking place via secular modes \citep{Youdin2011,Shariff2011,Takahashi2014}, the problem has significantly weakened. However, these theories require significant growth to Stokes numbers, $St = t_s \Omega$, of order 1 or 0.1 (where $t_s$ is the particle's stopping time and $\Omega$ is the dynamical time-scale) \reply{and mid-plane dust-to-gas ratios of order unity, typically requiring weak turbulence}. Since it is precisely these sizes that drift most rapidly towards the star, there remain open questions about whether growth into the required sizes can be efficient enough.

In addition to fragmentation, growth may be slowed by the tendency of large grains to bounce off each other  at moderately low velocity (a few $\mathrm{cm\,s}^{-1}$), which is most relevant for compact (i.e. low porosity) grains  \citep{Seizinger2013a}. For the smallest grains, which are very porous, bouncing is not a problem and growth proceeds efficiently. However, as grains become larger they become more compact because the collision velocities and gas pressure increase \reply{\citep[e.g.][]{Kataoka2013}}. Thus bouncing may become important at mm sizes \citep{Ormel2007a,Zsom2008,Ormel2009}. As well as porous grains being less susceptible to bouncing, the fragmentation threshold is also higher ($\sim 10 \unit{m\,s}^{-1}$, \citealt{Meru2013b}). Beyond the snowline where grains are icy, the increased stickiness suggests that even mm-sized grains remain porous enough that bouncing is not important \citep{Okuzumi2009,Okuzumi2012,Krijt2015}.

For compact grains bouncing at mm-size remains a potential barrier to growth. However, \citet{Windmark2012a} showed that by maintaining a population of small grains at the bouncing barrier, if a few particles manage to grow beyond the bouncing barrier then they may grow rapidly by sweeping up the smaller particles. This process relies on the fact that collisions between particles of very different masses can lead to another mode of growth at high velocity. This process, known as mass transfer, occurs because partial disruption of one of the grains can dissipate the collisional energy (\citealt{Teiser2009,Kothe2010}, see also \citealt{Guttler2010} for a collation of experimental results). A similar idea for pebble--planetesimal collisions (although made possible by the planetesimal's gravity rather than material strength), provides a promising formation mechanism for the cores of Jupiter-mass planets \citep{Ormel2010,Lambrechts2012,Morbidelli2012,Chambers2016}. 

Models of coagulation that take into account the full distribution of collision velocities suggest a possible way in which a small fraction of particles may grow beyond the bouncing barrier \citep{Windmark2012b,Garaud2013,Drazkowska2014,Estrada2016}. The key point is that due to turbulence there is a broad distribution of collision velocities between any pair of particles of a given size. Thus while the majority of collisions will occur near the peak of the distribution the vast number of dust grains means that rare occurrences of collisions at velocities in the tail of the distribution are almost guaranteed to occur occasionally. Thus some particles will be lucky enough to grow by a series of low velocity collisions, perhaps far enough that they enter the sweep up region of growth.

Indeed, such breakthrough has been observed in a number of models \citep{Windmark2012b,Garaud2013,Meru2013a,Drazkowska2014,Estrada2016}. In addition to lucky growth these models also have the advantage that they predict that the population of small grains is replenished by occasional high velocity collisions even when the average collision velocity is below the fragmentation threshold. However, introducing the full range of collision velocities has also had the side effect of making the growth considerably more difficult to follow numerically as we are interested in rare events. In models that trace the evolution of material between fixed mass bins, it is therefore possible for numerical diffusion across mass bins to swamp the growth rate due to lucky collisions. \citet{Drazkowska2014} demonstrate that this can lead to a massive overestimate of the rate of particles breaking through the bouncing barrier, and an underestimate of the time at which it happens.\footnote{Monte-Carlo methods tend to have the opposite problem, i.e. due to their limited dynamic range in density they tend to overestimate the time to breakthrough. See \autoref{Sec:Methods}.}

\reply{Here we aim to quantify the uncertainty in the level of breakthrough found in previous works by using a numerical method that converges rapidly with resolution. We focus on local models of grain growth using simple prescriptions for the grain micro-physics to allow a direct comparison to early works \citep{Windmark2012b,Garaud2013}, but have also included a more realistic model akin to \citet{Windmark2012a}. Inevitably, models that treat the entire disc and include detailed micro-physics \citep[such as][]{Estrada2016} are needed to capture the full complexity of growth and radial drift.} In \autoref{Sec:Methods} we present the numerical methods used, \autoref{Sec:Kernel} we describe the model kernel and \autoref{Sec:Tests} we present a variety of tests to demonstrate the performance of the code. In \autoref{Sec:Breakthrough} we present our results on the conditions for breakthrough, and in \autoref{Sec:Discuss} and \autoref{Sec:Conclude} we present our discussion and conclusions.

\section{Numerical methods}
\label{Sec:Methods}

Numerical approaches to grain growth can be broadly split into two methods: either by solving the \citet{Smoluchowski1916} equation for the evolution of the number density of grains of a given mass, $n(m)$, on a mass-mesh \citep[e.g.][]{Kovetz1969,Spaute1991,Lee2000,Brauer2008,Garaud2013}, or via a Monte Carlo approach in which the outcome of individual collisions between representative particles is decided by randomly sampling the probability distribution \citep{Gillespie1975,Ormel2008,Zsom2008}\footnote{Other approaches do exist, e.g. \citet{Wetherill1990,Inaba1999}}. \citet{Drazkowska2014} showed that both approaches give reasonable agreement when breakthrough is likely. However, although Monte-Carlo methods do not tend to suffer from the diffusive growth discussed above since individual masses can be tracked, they struggle to achieve the dynamic range in number density needed to investigate breakthrough in less favourable conditions. We therefore choose to use the Smoluchowski equation approach  with a method designed to reduce diffusion \citep{Lee2000}, and thus solve the equation
\begin{align}
&\deriv{n(m)}{t} = \quad \frac{1}{2}  \int_0^m K(m_1, m-m_1) n(m_1) n(m-m_1) \diff m_1 \nonumber \\
&\quad+ \frac{1}{2}\iint_{0}^{\infty}F(m_1, m_2)n(m_1)n(m_2)p(m_1, m_2, m)\diff m_1\diff m_2\nonumber \\
&\quad - n(m) \int_0^\infty \left[K(m,m_1) + F(m,m_1)\right] n(m_1) \diff m_1   
\label{Eqn:Smoluchowski}
\end{align}
directly on a `mass-mesh'. The first two terms on the right hand side represent the formation rate of particles of mass, $m$, by: 1) two particles of mass $m_1$, and $m-m_1$  merging to form a particle of mass $m$ (coagulation) and 2) the fragmentation products of mass, $m$, produced by fragmenting collisions between particles of mass $m_1$ and $m_2$. The third term describes the rate at which particles of mass $m$ are removed through collisions with particles of all sizes that result in either coagulation or fragmentation. Here $n(m)$ is the number density of grains per unit mass and per unit volume, $K$ and $F$ are the coagulation and fragmentation kernels and $p(m_1,m_2,m)$ is the fraction of fragments that have mass, $m$, produced in a fragmenting collision between two particles of mass $m_1$ and $m_2$. Mass-conservation requires $p(m_1, m_2, m) = 0$ for $m > m_1 + m_2$ along with the normalization
\begin{equation}
\int_{0}^{\infty} p(m_1, m_2, m)\diff m = 1.
\end{equation}
\reply{Since collisions resulting in fragments with mass $m > \max(m_1, m_2)$ would better be considered some form of growth, we take $p(m_1, m_2, m) = 0$ for $m > \max(m_1, m_2)$. This avoids any excessive growth due to numerical diffusion in the fragmentation routine that could arise if fragments with mass above the largest particle size were included.}

When considering the problem of breakthrough, the Smoluchowski equation has typically been solved by discretizing the problem at fixed particle masses. However, as discovered by \citet{Ohtsuki1990} and discussed by \citet{Garaud2013} and \citet{Drazkowska2014} the approach used can lead to an artificial acceleration of the growth rate unless the resolution is high, which \citet{Drazkowska2014} refer to as growth due to numerical diffusion. For this reason, we follow \citet{Lee2000} (which is closely based on \citealt{Spaute1991}) in taking into account the distribution within each mass bin. We will refer to these different approaches as `coarse' vs. `fine' as a result of their sub-grid interpretation. In \autoref{Sec:Tests} we will compare the two methods to simple cases and test problems with analytical solutions in order to demonstrate the improvement. First we describe both methods.

\subsection{Coarse sub-grid approximation}
This is the approach that has previously been applied to the breakthrough problem. In this approach \autoref{Eqn:Smoluchowski} is solved at discrete points on a mass-mesh $\left\{m_1, m_2, ..., m_I\right\}$ \citep{Podolak1980,Brauer2008,Garaud2013}.  \autoref{Eqn:Smoluchowski} is discretized in terms of the total number of grains (per unit volume) in a bin, $N_i$, of mass, $m_i$, resulting in the equation
\begin{align}
\deriv{N_i}{t} = 
    & \frac{1}{2} \sum_{j,k=1}^{I} C_{ijk} K_{jk} N_j N_k - \sum_{j=1}^I K_{ij} N_i N_j \nonumber \\
    &+\frac{1}{2} \sum_{j,k=1}^{I} F_{jk} N_j N_k p_{jki} - \sum_{j=1}^I F_{ij} N_i N_j.
\label{Eqn:Discrete}
\end{align}
We use the kernels evaluated at each point on the mass-mesh,  $K_{ij} \equiv K(m_i, m_j)$ and $F_{ij} \equiv F(m_i, m_j)$. We set the distribution of fragment products, $p_{ijk}$, to the integral of $p(m_i, m_j, m_k)$ over the bin to ensure $\sum_k p_{ijk} = 1$. Typically $p_{ijk}$ is chosen so that the number density of fragments follow a power-law with index $-\xi$ below some mass cut off.  We use $\xi = 11/6$, which is the standard choice \citep{Brauer2008,Garaud2013} based on the slope of interstellar extinction and arguments based on a self-similar collisional cascade \citep{Dohnanyi1969,Mathis1977,Draine1984,Tanaka1996}. 

The coagulation coefficients $C_{ijk}$ determine the fraction of the coagulation products that go into a given bin. For linear mass bins they are simply given by $C_{ijk} = \delta(j+k-i)$, but in order to resolve growth over many orders of magnitude logarithmic bin sizes are needed. Following \citet{Kovetz1969} and \citet{Brauer2008} we set  
\begin{equation}
 C_{ijk} = \begin{cases} f_{jk} &   i=i-, \\ 
                         1-f_{jk} & i=i+, \\
			 0 & \mathrm{otherwise},
           \end{cases}
\end{equation}
where the bin indices $i+$ and $i-$ are respectively the minimum and maximum indices that obey ${m_{i-} < m_j + m_k < m_{i+}}$ and
\begin{equation}
 f_{jk} = \frac{m_{i+} - (m_j + m_k)}{m_{i+} - m_{i-}}.
\end{equation}
These expressions can be derived by splitting the merger products between the two bins in a way that conserves both the mass and total collision rate, $K_{jk} N_j N_k$. An analogous way of thinking about these coefficients is that the merger products are assumed to be uniformly distributed between the limits $(m_j + m_k) \pm \frac{1}{2}(m_{i+} - m_{i-})$ before being assigned to their nearest point on the mass mesh.

To see why this scheme can give rise to a shorter growth time-scale, consider the growth of a grain of mass $m_k$ that grows by collisions with grains of mass $m_k \ll m_j$, such that the final product falls in the range $[m_j, m_{j+1}]$. For $m_k \ll m_j$ many collisions may be required to reach the mass $m_{j+1}$, but the algorithm above will place some mass in bin $j+1$ even before particles have had the time to grow to that size. We note that since there is a distribution of sizes within the mass-bin, there should be some growth into the largest mass-bin from particles near the upper end of the mass-range. However, whether or not these coefficients produce an accurate estimate of the growth rate from this small part of the bin depends sensitively on the distribution of densities. Thus in the steeply declining tails these coefficients will over-predict the growth rate.

Rather than evolve $N_i$ directly we use the mass in each bin $N_i m_i$ as the fundamental variable and instead use \autoref{Eqn:Discrete} to compute the rate of growth from each bin. In doing so we explicitly ensure that mass is conserved. Since these coefficients place a constant fraction of the coagulation product in the upper size bin independently of the mass distribution this explains why the coarse method produces excessive diffusion in the tails of the mass-distribution. \citet{Drazkowska2014} suggest that this can be controlled by suppressing growth in bins with very low density, and indeed they found it can help reduce spurious breakthrough at low resolution. We have not applied such a suppression to allow a more direct comparison to the fine sub-grid approximation, in which it is not needed.

\subsection{Fine sub-grid approximation}
Here we use an approach that closely follows \citet{Lee2000}. First, we assume the mass distribution within the $k$-th mass bin is given by a power-law, 
\begin{equation}
 \rho_k(m) = m n_k(m) = c_k (m/m_{k-1/2})^{q_k},
\end{equation}
where we use central differences to evaluate the log-space slopes, $q_k$,:
\begin{equation}
 q_k = \frac{ \log\left( \frac{M_{k+1}}{\Delta m_{k+1}} / \frac{M_{k-1}}{\Delta m_{k-1}}\right)}{\log \left(m_{k+1} / m_{k-1}\right)}.
\end{equation}
where $\Delta m_k = m_k - m_{k-1}$. At the edges of the domain, or where either of $m_{k\pm1} = 0$ we use one-sided differences instead. If either of these differences are not available we instead use $q_{k-1}$, although in practice this is only relevant at the highest mass bin, where growth and fragmentation can cancel out producing zero densities. Finally $c_k$ can be set using the normalization $\int \rho_k(m) \diff m = M_k$,  the mass per unit volume of particles within bin $k$. 

The mass-loss rate from bin $j$ due to coagulation with particles from bin $k$ is given by
\begin{equation}
\iint K(m, m') \rho_j(m) \frac{\rho_k(m')}{m'} \diff m \diff m',
\end{equation}
and the integration is over the ranges ${m_{j-1/2} < m < m_{j+1/2}}$ and ${m_{k-1/2} < m' < m_{k+1/2}}$. Even for constant $K(m, m')$ over the range of interest this expression cannot be efficiently evaluated for two power-laws and instead we are forced to make simplifications. 

The first simplification we make is to neglect the distribution of densities for the lower mass size bin, instead taking ${\rho_j = M_j \delta(m-m_j)}$, where $\delta(x)$ is the Dirac-$\delta$ function and $j < k$\footnote{For $m_j \approx m_k$ this will halve the range of masses over which the coagulation products are distributed. However, in this case the merger products will be in some bin $i > \max(j,k)$, thus the mean growth is handled adequately. For particles with $m_j \ll m_k$ this expression becomes increasing accurate, preventing the artificial growth of a large amount of mass into the bin $k+1$.}. After this simplification the rate of growth into bin $i$ from bins $j$ and $k$ with $j<k$ is 
\begin{align}
N_j \int_{m_-}^{m_+} &K(m_j, m) \frac{\rho_k(m)}{m}\left[ m_j + m \right] \diff m \approx \nonumber \\
&\bar{K} N_j \int_{m_-}^{m_+} \frac{\rho_k(m)}{m} \left[ m + m_j\right] \diff m,
\end{align}
where $N_j = M_j / m_j$ is the number of grains in bin $j$. \reply{The second simplification} is that we make an approximation in evaluating the kernel using a weighted sum:
\begin{equation}
 \bar{K} = \frac{K(m_-, m_j) w_- + K(m_+, m_j) w_+}{w_- + w_+},
\end{equation}
where $w_{\pm} = \rho_k(m_\pm)(m_\pm + m_j)/m_\pm$. The integration limits are given by the overlap of the mass range of the coagulation product $[m_j + m_{k-1/2}, m_j + m_{k+1/2}]$ with the mass range of bin $i$, $[m_{i-1/2}, m_{i+1/2}]$, which gives 
\begin{align}
m_- &= \max(m_{k-1/2}, m_{i-1/2} - m_j), \\
m_+ &= \min(m_{k+1/2}, m_{i+1/2} - m_j).
\end{align}
The mass-loss from bins $j$ and $k$ is computed using the same weighted average to ensure mass conservation.

The fragmentation calculation could be extended similarly by using weighted averages of the fragmentation kernel, $F_{ijk}$. However, treating fragmentation carefully is much less important than treating coagulation carefully because fragmentation cannot lead to diffusive mass growth and so the difference that it makes is small. Instead we simply use the same approach to fragmentation as for the coarse sub-grid approximation.

In the fine sub-grid approximation we have also implemented the active bins method of \citet{Lee2000}, in which the computation of the coagulation and fragmentation rates from bins below a threshold density is skipped. This is necessary since the power-law slope is ill-defined when the bins are empty. To ensure  accuracy, we include the contribution from all bins that have mass above the minimum value that can be represented in double precision.

\subsection{Time Integration}

We have investigated three different approaches to time-stepping, including an explicit 2nd order Runge-Kutta.
Due to the inherent stiffness of the coagulation-fragmentation equations we find that using a semi-implicit time integration results in a significant speed up. For most purposes a semi-implicit time-centred Euler integration is adequate. Rather than compute the full Jacobian, we find that it is both sufficient and faster to use an approximate Jacobian when using the fine sub-grid model. We do so by neglecting the contribution to the Jacobian that arises from the variation of $q_k$. When using either the Runge-Kutta or implicit Euler we find limiting the change in the solution to one per cent each time-step is usually sufficient to ensure stability, with excellent agreement between both methods. However, in some simulations the tail of the distribution became unstable on approach to equilibrium when using implicit Euler integration. While this could be prevented by setting an upper limit to the time-step, this can be prohibitively expensive when exploring a grid of models as it requires tuning by hand. A practical alternative is to use a semi-implicit Rosenbrock method with embedded step-size control \citep[see, e.g.][]{Press2002}. We choose the 3rd order Rosenbrock method of \citet{Rang2005}, which is stable when using an approximate Jacobian. This is the scheme used for the models presented in this paper.

\subsection{Relative computational cost of the methods}

\reply{We have not conducted an extensive comparison between computational efficiency of the methods; however, we note that despite the higher computational cost per mass bin of the fine-sub grid approximation it can still be competitive. In pure coagulation tests with 150 to 350 mass bins the fine sub-grid model is 30 -- 50 times slower per cell (when the active bins optimization is not used). However, since a greater accuracy can be expected for the same number of bins, the computational time need to achieve the same desired accuracy can be considerably less. The difference in the number of bins required will be application dependent, but as an example a factor of $\sim 3$ fewer bins are needed for the same accuracy in the bulk of the mass distribution for the linear kernel test in \autoref{Sec:Tests}, reducing the difference in cost to a factor of 3 -- 6 for the same accuracy (due to the $O(N^2)$ scaling). However, for problems where one is interested in the mass distribution in the tails of the distribution, as we are here, the coarse sub-grid model requires a far higher resolution (in fact impractically high) to achieve the same accuracy  (see \autoref{Fig:MTC} and \autoref{Fig:phys_compare}), preventing it from being competitive.}

\reply{When fragmentation is included, normally the difference in computational cost for the same number of bins becomes negligible at high resolutions because the fragmentation calculation scales as $O(N^3)$, where N is the number of bins, while the coagulation routines scale more favourably, $O(N^2)$. In this case, the requirement of fewer bins for the same accuracy can mean that fine sub-grid approximation will produce an accurate solution in a shorter absolute time too. This is our experience for the more `realistic' problems discussed in \autoref{Sec:Breakthrough}.}

\reply{As a further note, under certain approximations for $p(m, m_1, m_2)$ the fragmentation rate can actually be computed in $O(N^2)$ when explicit time integration is used. One such case is when the approximation $p(m, m_1, m_2) \equiv p(m, \max(m_1, m_2))$ is made. When an $O(N^2)$ calculation of the fragmentation rate is possible, the additional cost of the fine-grid model may be more evident. However, the cost of the Jacobian calculation remains $O(N^3)$, thus implicit or semi-implicit methods such as the one we use here cannot benefit from this optimization, and thus the fine-grid model will remains the most expedient.}

\section{Model kernel}
\label{Sec:Kernel}

\reply{A variety of different model kernels have been used to study grain growth. In reality, the outcome of collisions between grains is extraordinarily complex, depending on the velocity and angle of impact and internal properties of both grains (see e.g. \citealt{Garaud2013} for a discussion) and thus it is necessary to simplify the problem. This is typically done by breaking the kernel down into two parts: a model for the distribution of collision velocities and a model for the outcome of the collisions. When studying lucky growth and breakthrough, typically a simple model for the collisional outcomes has been used \citep{Windmark2012b,Garaud2013,Drazkowska2014}. Although more complex models based upon comparison with experimental data exist \citep[e.g.][]{Guttler2010,Windmark2012a} these generally have not been combined with the full distribution of collision velocities required to study lucky growth, except in one global model \citep{Estrada2016}.}

\reply{Similarly, a large variety of approximations for the collision velocity have been used, varying from using just the mean \citep{Brauer2008,Birnstiel2010} to several different models for the full distribution of collision velocities \citep{Okuzumi2011,Windmark2012b,Garaud2013}. We follow \citet{Garaud2013} for the distribution of collision velocities between a pair of particles of arbitrary size, which represents the current state of the art.} For each grain size independently, we approximate the distribution of velocities in each direction as a Gaussian with the mean given by the motion relative to the gas and standard deviation given by the random motions induced by Brownian motion and turbulence. The combined collision velocity distribution is then computed by \reply{assuming the velocity distributions are independent and} integrating over the distribution, which \citet{Garaud2013} show is given by:
\begin{align}
P(\Delta_{ij}) = \frac{1}{\sqrt{2\upi} \sigma_{ij}} \frac{\Delta_{ij}}{\bar{\Delta}^D_{ij}}
& \left[        \exp \left(- \frac{(\Delta_{ij} - \bar{\Delta}^D_{ij})^2}{2 \sigma_{ij}}\right)\right. \nonumber\\
& \left. \quad -\exp \left(- \frac{(\Delta_{ij} + \bar{\Delta}^D_{ij})^2}{2 \sigma_{ij}}\right)\right] ,
\label{Eqn:CollDist}
\end{align}
where $\Delta_{ij}$ is the velocity of the collision, $\bar{\Delta}^D_{ij}$ is the non-random velocity differences between the two sizes (e.g. difference in radial drift velocities) and $\sigma_{ij}$ is the r.m.s. turbulent velocity. For more details see \citet{Garaud2013}. 

The extent to which the underlying \reply{single particle} velocity distributions are well represented by Gaussian distributions is however still somewhat unclear -- analytic theories are only able to compute the r.m.s collision velocity and not the whole distribution \citep{Volk1980,Zaichik2003,Ormel2007,Pan2010}, while direct numerical simulations are typically limited to low Reynolds number, $R_{\rm e}$ typically not exceeding $\sim500$, which inhibits their ability to probe the inertial range scaling that is important for high $R_{\rm e}$ applications \citep[e.g.][]{Carballido2010,Nelson2010, Pan2013,Pan2014,Ireland2016}. Shell models of turbulence are able to overcome the limitation of low $R_{\rm e}$ but do not capture the \reply{non-Gaussian behaviour} in the statistics of the underlying turbulence \citep{Hubbard2012, Hubbard2013}. \reply{Furthermore, the simulations are generally conducted for homogeneous turbulence and thus we can only test the special case of $\bar{\Delta}^D_{ij}=0$ directly. In this case $P(\Delta_{ij})$ reduces to a Maxwellian \citep{Garaud2013}.}

However, the simulation results are promising. At least for particles that are well within the inertial range and differ in size by more than a factor of a few the velocities are only weakly correlated and the \reply{single particle} velocity distributions are well approximated by a Gaussian. Indeed, \autoref{Eqn:CollDist} is in excellent agreement with the distribution of collision velocities for particles with size ratios greater than about 2, at least within the inertial range \citep{Hubbard2013}. Similarly, for particles with stopping times longer than the turnover time of the largest eddies ($St > 1$), a Gaussian distribution is again an excellent approximation and thus once again \autoref{Eqn:CollDist} is an excellent model for the collision velocities \citep{Pan2013,Ireland2016}. While particles in the dissipation range of the turbulence do exhibit correlated behaviour, for reasonable choices of parameters the particle sizes that we are interested in ($>10\unit{\umu m}$) are well within the inertial range. \reply{Thus the only case in which any deviation from \autoref{Eqn:CollDist} may be significant is for collisions between similar sized particles with $\St < 1$.} 

\reply{For particles with similar sizes and $\St < 1$, correlated motions lead to lower collision velocities on average, but broader tails than predicted by \autoref{Eqn:CollDist}. Since away from any growth barriers the growth rate is normally dominated by collisions between similar sized particles, this effect could make a difference to the growth rate. However, away from the bouncing barrier the growth rate is controlled by the collisions occurring at the peak of the distribution and the wider tails are less significant. The reduction in peak collision velocity associated with the correlated motions is taken into account by using the pairwise r.m.s turbulent velocities from \citet{Ormel2007}, which include this effect, at least approximately\footnote{See \citet{Pan2010, Pan2015}, for example, for criticisms of their approach.}. Close to the bouncing barrier collisions in the tail of the distribution become more important, but growth also becomes dependent on collisions between particles of different sizes. Thus the wider tails for similar sized particles may not be of critical importance, despite their contribution to the growth and fragmentation rates being underestimated.}

For the background relative velocity we again use the disc model from \citet{Garaud2013}, which is set by radial drift, azimuthal drift, and settling:
\begin{align}
v_{r, i}      =& \frac{1}{1 + \mathrm{St}_i^2}(v_{r,\mathrm{g}} - 2 \mathrm{St}_i \eta v_\mathrm{K}) \\
v_{\theta, i} =& \frac{1}{2 \mathrm{St}_i} (v_{r,i} - v_{r,\mathrm{g}}) \\
v_{z, i}      =& - h_i \Omega_\mathrm{K} \min\left(\mathrm{St}_i, \frac{1}{2} \right).
\end{align}
The Keplerian velocity, angular frequency, and Keplerian decrement are $v_\mathrm{K}$, $\Omega_\mathrm{K}$, and $\eta = - \frac{r}{2v_{\rm K}^2}\frac{1}{\rho_g}\deriv{P}{r}$, while the gas radial velocity, $v_{r,\mathrm{g}}$, is taken to be the viscous velocity. $\mathrm{St}_i$ is the Stokes number of the particle and $h_i$ is scale-height of the dust layer in the disc. The dust scale height is set via
\begin{equation}
  \frac{1}{h_i^2} = \mathrm{St}_i(1 + \mathrm{St}_i) \Omega_k \frac{\mathrm{Sc}}{\nu} + \frac{1}{H^2},
\end{equation}
where $\nu$ is the turbulent viscosity, $\mathrm{Sc}=1$ is the Schmidt number (ratio of viscosity to mass diffusivity) and $H$ is the gas scale-height \citep{Dubrulle1995, Garaud2007}. We assume an alpha model for the turbulent viscosity with the default choice being $\alpha = 10^{-4}$. The Stokes number is set assuming Epstein drag and spherical particles, $\mathrm{St}_i = t_s \Omega_K = \Omega_K s_i \rho_\mathrm{s} / \rho_\mathrm{g} c_\mathrm{s}$, where $\rho_\mathrm{s,g}$ are the solid and gas densities at the mid plane and $c_\mathrm{s}$ is the local sound speed. Finally we include the contributions to the velocity dispersion from turbulence and Brownian motion (added in quadrature).

The model coagulation and fragmentation kernels, which directly take into account the distribution of relative velocities, can be written as
\begin{align}
 K_{ij} =& \bar{a}_{ij} \int_0^\infty \Delta v P_{ij}(\Delta v) \epsilon^s_{ij}(\Delta v) \diff \Delta v \\
 F_{ij} =& \bar{a}_{ij} \int_0^\infty \Delta v P_{ij}(\Delta v) \epsilon^f_{ij}(\Delta v) \diff \Delta v, 
\end{align}
where $\bar{a}_{ij} = \pi (s_i + s_j)^2$ is the collisional cross-section of two spherical particles of size $s_i$ and $s_j$, and $P_{ij}(\Delta v)$ is the distribution of collision velocities between two particle sizes. The factors $\epsilon^{s,f}_{ij}$ are the probabilities that collisions at a given velocity result in coagulation or fragmentation, which differ between the two models below.

\begin{table}
\caption{Default model parameters used for defining the coagulation and fragmentation kernels}
\begin{tabular}{lccl}
\hline
Symbol & Value & Units & Description \\
\hline
$\alpha$    &  $10^{-4}$ &                    & Turbulence parameter  \\
$v_{\rm b}$ &  5         & ${\rm cm\,s}^{-1}$ & Bouncing threshold \\
$v_{\rm f}$ &  100       & ${\rm cm\,s}^{-1}$ & Fragmentation threshold \\
$\phi$      &  100       &                    & Critical mass transfer ratio\\
\hline
\end{tabular}
\label{Tab:ModelParam}
\end{table}

\begin{figure*}
 \centering
\begin{tabular}{cc}
\includegraphics[width=\columnwidth]{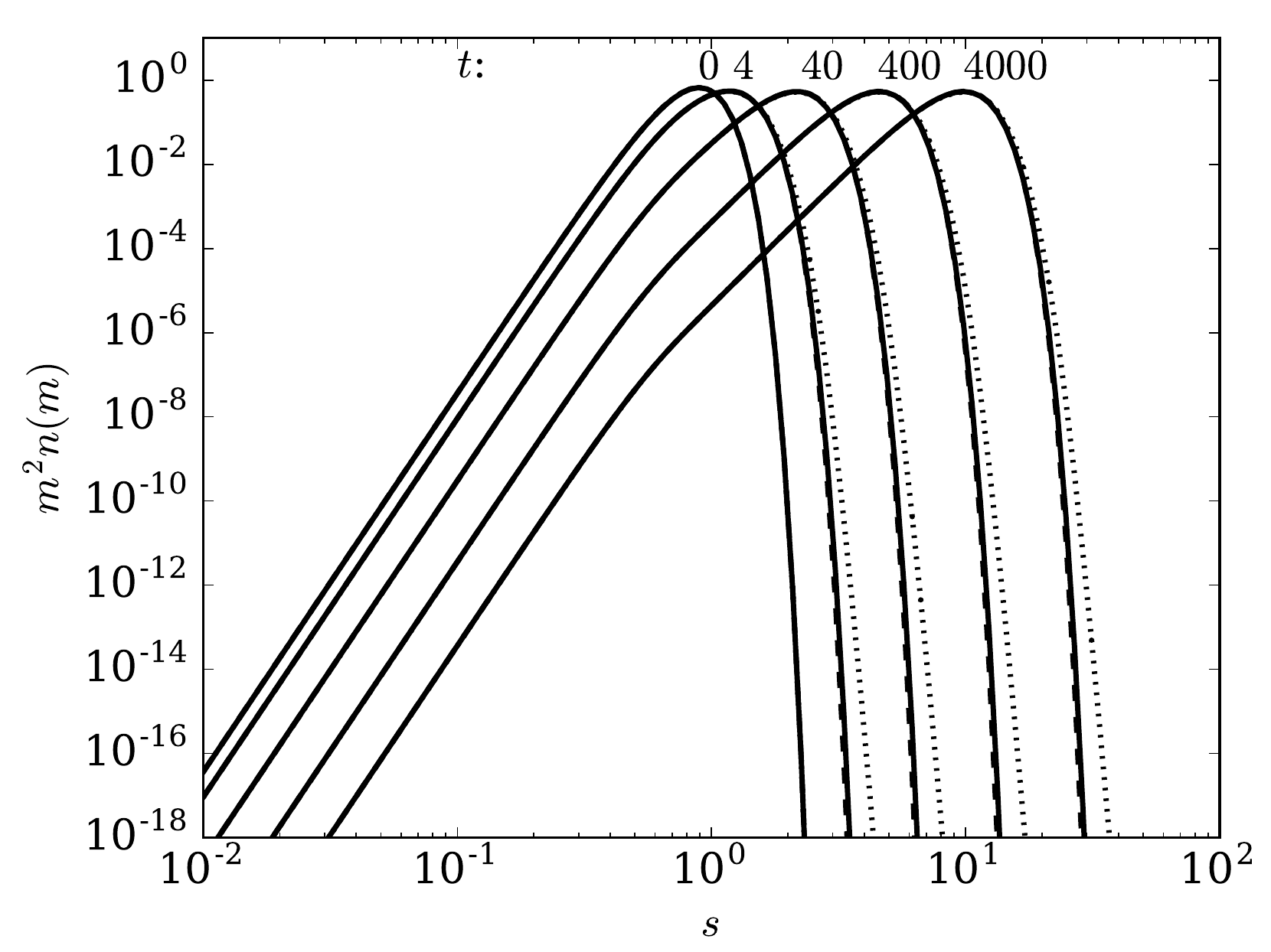} &
\includegraphics[width=\columnwidth]{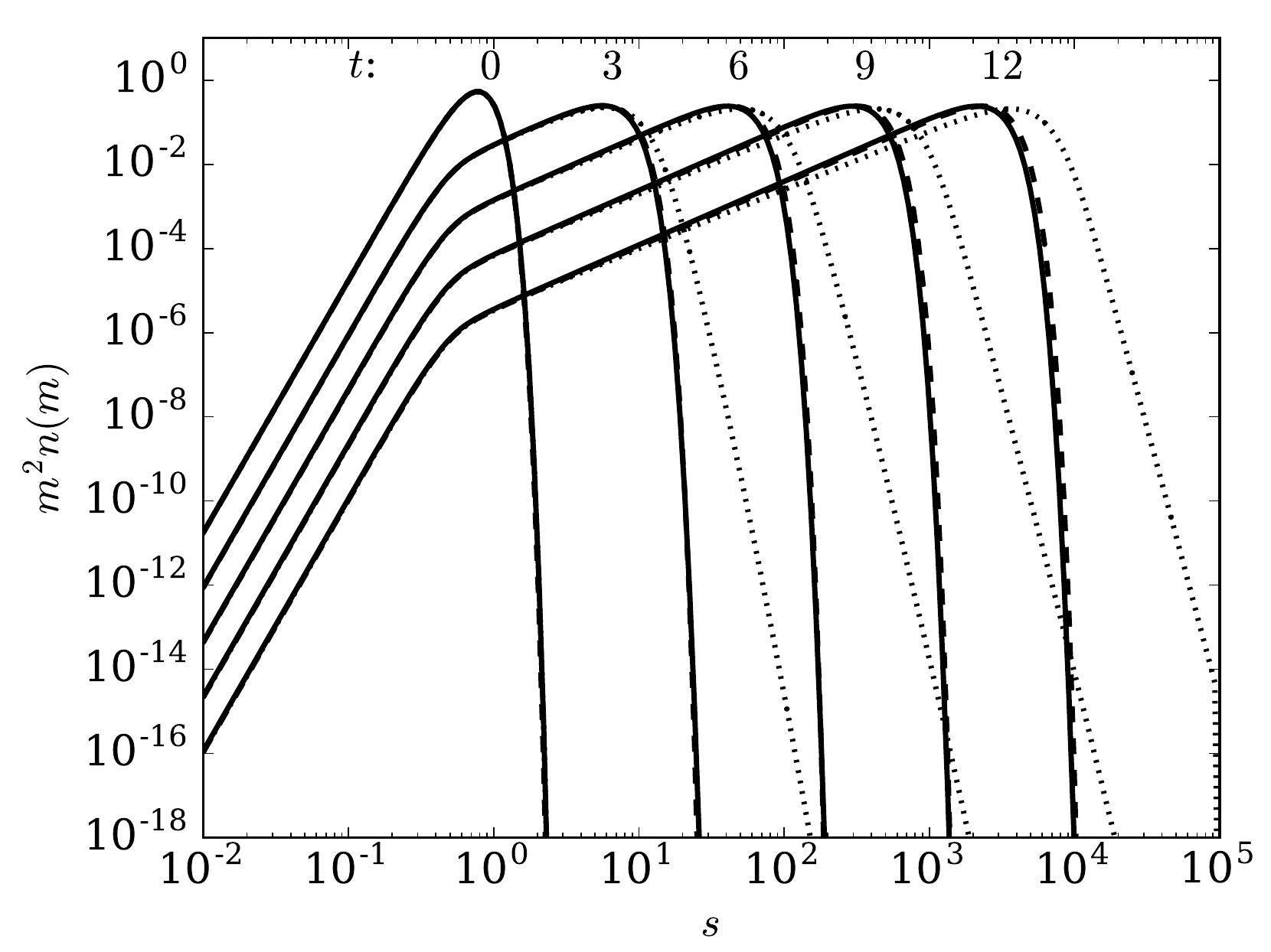} 
\end{tabular}
\caption{Growth in pure coagulation models for simple kernels, $K(m_i, m_j) = 1$ (left) and $K(m_i, m_j) = m_i + m_j$ (right). The analytical solutions are shown as the dashed black lines, along with numerical solutions using 10 bins per logarithmic interval in mass for the fine (solid) and coarse (dotted) sub-grid models. Units are dimensionless. The fine sub-grid approximation reproduces the analytical solution more accurately, particularly in the tails (the solid and dashed lines lie almost on top of each other).}
\label{Fig:SimpleKern}
\end{figure*}

\reply{To complete the model kernel, we need a prescription for the sticking and fragmentation probabilities, $\epsilon^{\rm s,f}$, along with a model for mass transfer. Previous, detailed studies of lucky growth and breakthrough used a simple model for the kernels where the outcome of a collision depends only on the velocity with which the particles collide \citep{Windmark2012b,Garaud2013,Drazkowska2014}. In this simple model, collisions with a velocity below a critical value, $v_{\rm b}$, result in sticking ($\epsilon^s = 1$) and above this bouncing occurs instead ($\epsilon^s = 0$). Collisions at velocities above  a further threshold, $v_{\rm f}$, lead to fragmentation of the grains ($\epsilon^f = 1$). In the above works, mass transfer is treated by assuming that at mass ratios above a certain threshold, $\phi$,  high velocity collisions that would lead to fragmentation instead lead to growth via mass transfer. \citet{Windmark2012b} and \citet{Drazkowska2014} allowed for mass transfer to be inefficient, i.e. only a small fraction (of order 10 per cent) of the smaller particle is transferred to the larger particle with the rest of the mass being distributed as small fragments. Here we have assumed that all of the mass is transferred. While this model of mass transfer is overly optimistic, we will demonstrate that the simple model may actually under predict lucky growth due to over estimating the fragmentation rate. The default parameters used in this model are summarized in \autoref{Tab:ModelParam}.}

\reply{In addition to the simple model of collisional outcomes, we also explore a second, more physically motivated model. Our motivation is that the simple model of a constant threshold velocity for fragmentation over-estimates the fragmentation rate. To see this, consider first a collision between two equal mass particles at a velocity just high enough to cause fragmentation. By considering the energy available in the collision, we should expect that if one of the particles is replaced by one much less massive, then collisions at the same velocity as before should not have enough energy to disrupt the larger particle. Thus the threshold fragmentation velocity should at least depend on the mass ratio of the particle pair. This idea is borne out by comparison with experiment \citep{Stewart2009,Beitz2011,Windmark2012a} and supported by numerical simulations \citep[e.g.][]{Meru2013a}. The reason that the mass-ratio dependence of the fragmentation threshold is important is that the relative velocity between the largest particles and small particles is usually larger than the velocity between two of the largest size particles in both the turbulent and drift dominated regimes. Therefore, as particles enter the regime where fragmentation is important, the numerous collisions with small particles (which are much more abundant) always dominate the fragmentation rate if a constant fragmentation threshold is assumed. Not only does this mean the fragmentation rate is over-estimated, it is also sensitive to the minimum size of particles present as they are the most abundant.}

\reply{A more realistic model of fragmentation was presented by \citet{Windmark2012a} through fits to experimental data \citep{Beitz2011}. The fits suggested by \citet{Windmark2012a} are in remarkable agreement with the physically motivated model of \citet{Stewart2009}, which is based on the comparison of the specific collision energy to the fragmentation threshold. Given our simple model of bouncing, and the considerable uncertainty still inherent in the experimental data, we instead adopt a simple model intended to capture the scaling of the fragmentation threshold with mass ratio. \citet{Leinhardt2012} show that for a fixed target mass, the threshold should scale as}
\begin{equation}
v_{\rm f} = v_{{\rm f}, \phi=1} \left(\frac{(1+\phi)^2}{4\phi}\right)^{1/(3\mu)},
\end{equation}
\reply{where $v_{{\rm f}, \phi=1}$ is the threshold for an equal mass collision and $\phi \ge 1$ is the mass ratio. The parameter $\mu$ depends on the material properties, and is in the range $1/3 < \mu < 2/3$. Both the fits suggested by \citet{Windmark2012a} and the simulations of \citet{Leinhardt2012} are well approximated by $\mu = 1/3$, and so we adopt $\mu = 1/3$ here. Strictly, the  equal mass fragmentation threshold is a weak function of the grain mass as well. However, we keep $v_{{\rm f}, \phi=1}$ fixed to the same value as the simple model, so that it is clear that the differences between the two models arise due to differences in the way that fragmentation driven by small grains is treated.}

\reply{The above expression is appropriate for total fragmentation, however at large mass ratio the small grain may fragment at much lower velocity than the larger grain. For collision velocities between these thresholds this results in the fragmentation of the smaller grain, followed by partial re-accretion, is what drives mass transfer \citep[e.g.][]{Guttler2010}. To estimate the fragmentation threshold for the smaller grain, we assume that half of the collision energy may be available to fragment the smaller grain. Thus it fragments when the collision energy exceeds twice the fragmentation threshold energy for that grain. The resulting prescription for the fragmentation thresholds of the smaller and larger grains is:}
\begin{align}
v_{\rm f, s} =& v_{\rm f} \frac{1}{2} \left(1 + \phi^{-1} \right), \nonumber \\
v_{\rm f, l} =& v_{\rm f} \frac{(1 + \phi)^2}{4 \phi}. 
 \label{Eqn:FragThresh}
\end{align}
\reply{Comparing these expressions to \citet{Windmark2012a}, we see that fragmentation threshold for the smaller grain has the same form, but our expression for the fragmentation threshold of the larger grain is more conservative than theirs, $v_{\rm f, l} = v_{\rm f} (1 + \phi)/2$, allowing fragmentation at half the velocity for high mass ratios.}

\reply{The two different fragmentation thresholds above naturally suggest an improvement to the mass transfer model, which occurs when the collision velocity, $\Delta v$,  obeys $v_{\rm f,s} > \Delta v > v_{\rm f,l}$. In this regime some of the mass of the smaller particle may be transferred to the larger particle. For $\Delta v > v_{\rm f,l}$ we assume fragmentation of both grains, while $\Delta v < v_{\rm f,s}$ results in sticking or bouncing. \citet{Windmark2012a} use fits to experimental data to determine how much of the smaller particle is accreted. For simplicity, we assume here that the amount of mass transferred is given by some constant fraction of smaller grain's mass, $\epsilon^{\rm m}$,  for which we explore different values in the range $ 0 \le \epsilon^{\rm m} \le 1$. For simplicity, we modelled $\epsilon^{\rm m} < 1$ by treating the collisions as complete coagulation but at a reduced rate, rather than by reducing the amount of mass transferred. This gives approximately the correct growth rate, but neglects the production of small fragments in the collisions. However, this effect should be secondary to the reduction in the fragmentation rate and growth rate by mass-transfer.}

To close the system of equations we assume the same disc model as \citet{Garaud2013},  an exponentially tapered surface density, 
\begin{equation}
 \Sigma = \frac{M_{\rm disc}}{2 \upi r R_0} \exp (- r / R_0), \label{Eqn:SurfDens}
\end{equation}
with a Gaussian vertical structure. We assume that the sound speed obeys a power law $c_s \propto r^{-1/4}$. Finally, we use $M_{\rm disc} = 0.0375 M_\odot$, $R_0 = 30\unit{au}$ and additionally set $c_s = 1\unit{km\,s}^{-1}$ at $1\unit{au}$ along with a stellar mass of $0.75\,M_\odot$.

\section{Tests}
\label{Sec:Tests}

To verify the methods we first compare them against the analytical solutions to pure coagulation models with $K(m_i, m_j) = 1$ and $K(m_i, m_j) = m_i + m_j$, which bound the asymptotic behaviour of the model kernel we use. We initialize the solutions using the analytical solution at $t=0$ \citep{Wetherill1990}. The evolution of the test problems is shown in \autoref{Fig:SimpleKern} for both methods described using a modest resolution of ten points per logarithmic decade in mass ($N_{\rm bd} = 10$). We see that both methods reproduce the main features of the analytical solutions well, but the coarse method gives rise to more material in the tails. The coarse method also overestimates the peak size of the particle distribution for $K(m_i, m_j) = m_i + m_j$,  a result of the numerical enhancement of the growth rate, as discussed in \autoref{Sec:Methods}. The larger mass at the peak of the distribution and the diffusion in the tails demonstrate why \citet{Drazkowska2014} saw particles breaking through the bouncing barrier earlier at low resolution.

\begin{figure}
 \centering
\includegraphics[width=\columnwidth]{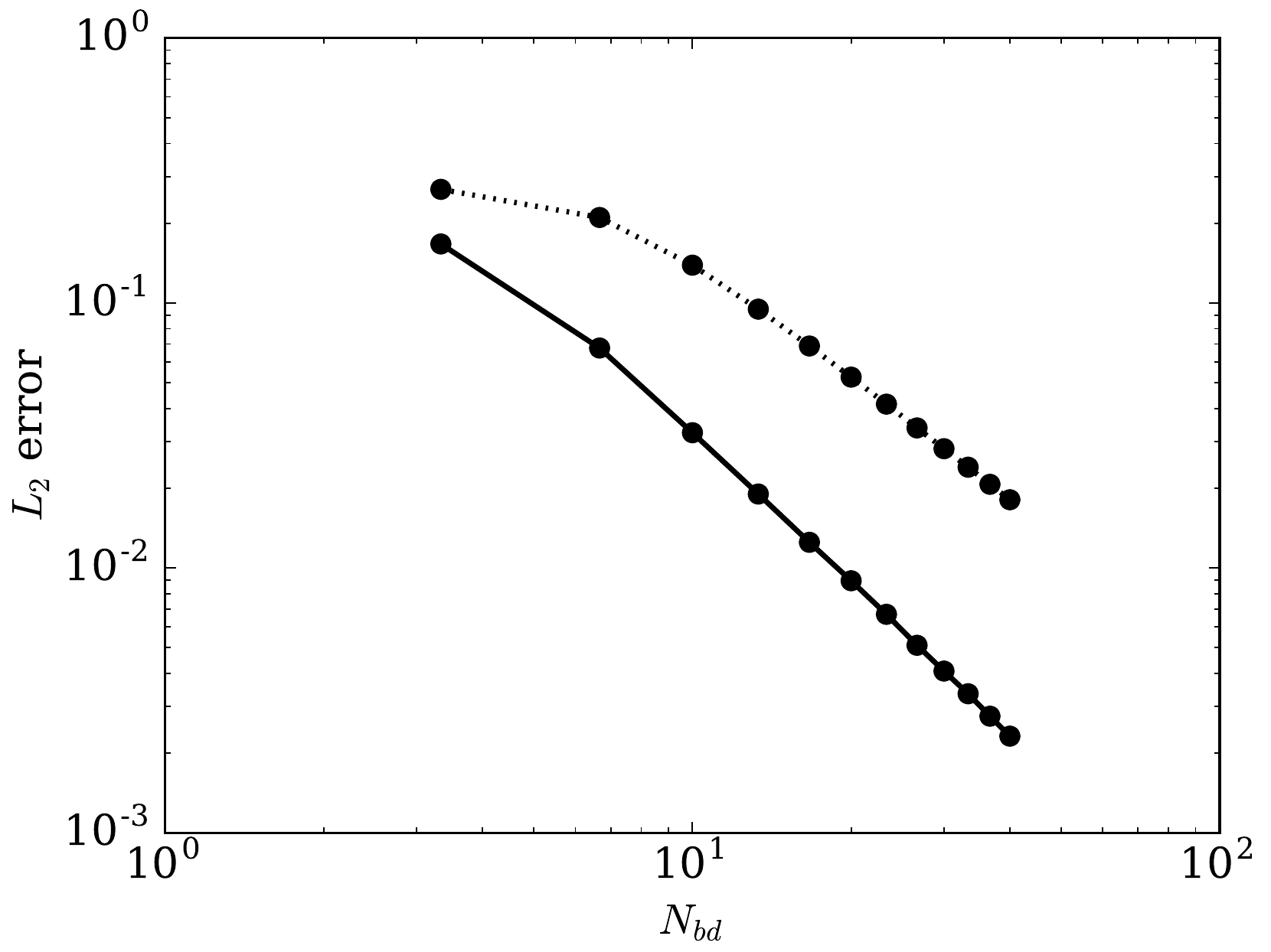} 
\caption{$L_2$ error norm for the fine (solid) and coarse (dotted) sub-grid models in a pure coagulation model with the linear kernel, $K(m_i, m_j) = (m_i + m_j)$. The $L_2$ error is computed for $m^2\tderiv{N}{m}$ at $t=12$ for different numbers of bins per mass decade, $N_{\rm bd}$.}
\label{Fig:ErrorLinear}
\end{figure}

To estimate the error in a more quantitative way, we use the $L_p$ error norm defined by
\begin{equation}
 L_p = \left(\frac{1}{N} \sum_i^N \left|\frac{f_i - f_{\rm exact}(m_i)}{\max(\{f_{\rm exact}(m_i)\})}\right|^p\right)^{1/p}
\end{equation}
where $f_i$ are the $N$ numerical values and $f_{\rm exact}(m_i)$ is the exact solution for the quantity $f$ evaluated at the same points, $m_i$, as the numerical solution. We normalize the error estimate to the maximum value of the analytical solution over the set of $m_i$. In \autoref{Fig:ErrorLinear} we show the $L_2$ error norm of the density per logarithmic interval in mass, $m^2\tderiv{N}{m}$, for the linear kernel $K(m_i, m_j) = (m_i + m_j)$. We see that as well as significantly less numerical diffusion, the fine sub-grid method shows lower error overall and a faster rate of convergence (close to second order). For $K(m_i, m_j) = 1$, the results are similar, but both methods show second order convergence, with the fine sub-grid method having lower overall error. We find similar results for the $L_1$ and $L_\infty$ (maximum deviation) \reply{error} norms.

\begin{figure}
 \centering
\includegraphics[width=\columnwidth]{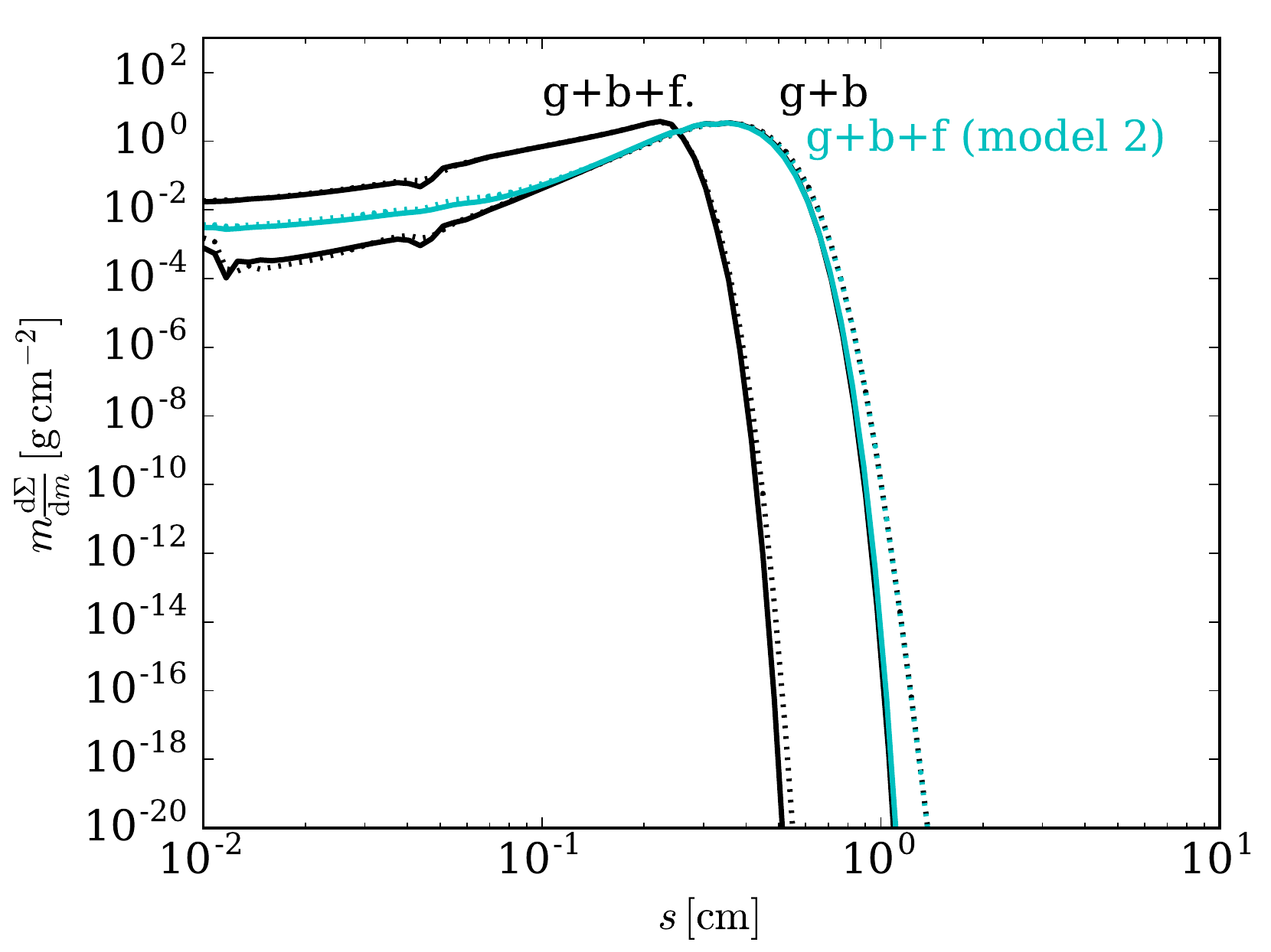} 
\caption{Particle distribution at $1\unit{au}$ after $30,000\unit{yr}$ for the growth+bouncing (g+b) and growth+bouncing+fragmentation (g+b+f) models. For the g+b+f model, the black line denotes a model in which the constant fragmentation threshold is used, while the cyan line shows the second model, which used the mass-ratio dependent threshold (\autoref{Eqn:FragThresh}), which produces a similar distribution to the g+b model with the constant threshold.  The different numerical methods are denoted by the line style, with the fine and coarse sub-grid models shown by the solid and dotted lines, respectively.}
\label{Fig:GBF}
\end{figure}

Since our model kernel lies somewhere between the two analytic cases it is useful to compare the methods on two simple problems using the model kernel. For this we choose a pure growth and bouncing model (g+b), along with a model in which we also include fragmentation \reply{according to the two prescriptions} (g+b+f). \reply{In the second, physically motivated fragmentation model, the larger fragmentation threshold (\autoref{Eqn:FragThresh}) was used for both the larger and smaller grains.} For the parameters, we use the default values in \autoref{Tab:ModelParam} evaluated at $1\unit{au}$. We set the minimum particle size to $0.1\unit{mm}$ and use $N_{\rm bd} = 10$ for this test. We assume the initial mass-distribution is a narrow distribution peaked at the size of the smallest particle, $0.1\unit{mm}$, which we represent as a Gaussian with width $10\unit{\umu m}$. \autoref{Fig:GBF} shows that both methods are generally in reasonable agreement with each other, with the coarse method still being somewhat more diffusive, but the difference in performance between the methods appears to be closer to the constant kernel case than the linear kernel.

\reply{Comparing the two different fragmentation models, we see that the maximum particle size is much smaller when a constant fragmentation threshold is used. Conversely, in the mass-dependent model based on \citet{Stewart2009} fragmentation only makes a small difference at this time, increasing the number of small grains but hardly affecting the maximum grain size. The difference highlights the role that small grains are playing in driving a fragmentation rate in the constant threshold model. In the g+b model and the model with the mass-ratio dependent fragmentation threshold the maximum particle size is still increasing slowly as the particles grow through the bouncing barrier. However, the high fragmentation rate in the constant fragmentation threshold g+b+f model has already driven the distribution into an equilibrium between growth and fragmentation. As noted by \citet{Garaud2013}, the unphysically high rate of fragmentation caused by small grains causes the maximum size of dust grains to be dependent on the size of the smallest grains used. This is not an issue for the mass-dependent fragmentation model, which we have confirmed by repeating the g+b+f model with a minimum grain size ten times smaller.}

The g+b and g+b+f models show that the two numerical methods are highly successful at capturing the bulk of the mass-distribution and along with the very steep cut off in the density distribution that makes resolving low levels of breakthrough so challenging. These tests show further evidence of excess diffusion in the tail of the coarse method, along with the slight shift in the peak density to larger sizes that is likely responsible for the strong resolution dependence on the breakthrough time found by \citet{Drazkowska2014}.

\section{Conditions for breakthrough}
\label{Sec:Breakthrough}
\subsection{Simple fragmentation model (fixed $v_{\rm f}$)}

\begin{figure}
 \centering
\includegraphics[width=\columnwidth]{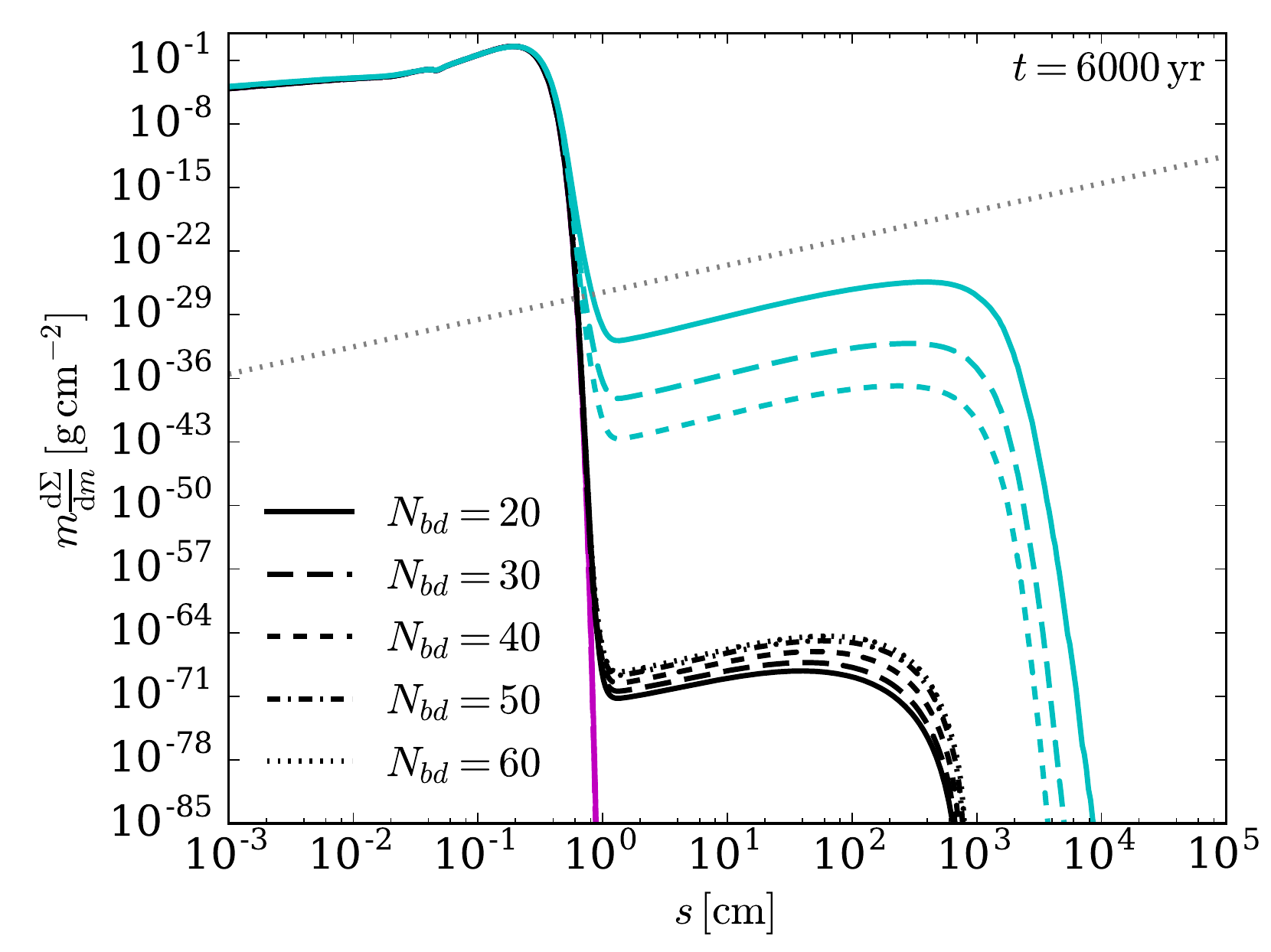} 
\caption{Particle distribution at $1\unit{au}$ for the full mass transfer model with the standard parameters computed at different resolutions. The blue and black lines show the solutions using the coarse and fine methods respectively. The magenta line shows a model where the mass transfer region of collisions results in bouncing rather than growth. The grey dashed line denotes one particle per logarithmic interval in mass and disc radius.}
\label{Fig:MTC}
\end{figure}

We first discuss the range of behaviours seen in the different models before exploring the parameter space in more detail. The typical behaviours vary from models that experience no breakthrough at all, to wholesale breakthrough along with models that show a low level of breakthrough in which a tiny fraction of particles are able to grow beyond the bouncing barrier. We base this discussion around models using our fiducial parameters  (\autoref{Tab:ModelParam}), considering the growth of particles at $1\unit{au}$ and varying the threshold for mass-transfer $\phi$. For all the models considered in this section we use a grid that extends from $1\unit{\umu m}$ to $10 \unit{km}$\footnote{Our coagulation model has not been designed for such large particles ($\sim\unit{km}$ size). Neglected processes, including the particles' gravity, are important for accurate modelling of these large sizes. For our purpose, which is to study breakthrough of the bouncing barrier, the detailed growth of these particles is less important than the fact that growth beyond the barrier at around mm-sizes can occur at all.}. For the initial conditions we assumed that the grains were all initially close to the minimum particle size used, $1\unit{\umu m}$. To ensure that the initial conditions were resolved we represent this with a  Gaussian distribution with a mean and standard deviation of $1\unit{\umu m}$ and $0.1\unit{\umu m}$ respectively, although the results are not sensitive to this choice. \autoref{Fig:MTC} shows the results of models with ${\phi=100}$ at $6000\unit{yr}$ computed with both methods. Instead of $n(m)$, we plot $m\deriv{\Sigma}{m} = m^2 n(m) H$, a proxy for the vertically integrated mass per unit logarithmic interval. In addition to the mass-transfer models, \autoref{Fig:MTC} includes a modified g+b+f model in which high velocity collisions (${v > v_f}$) with mass ratio above $\phi$ are assumed to result in bouncing instead of growth by mass transfer.

\reply{The main features of this model have been described in detail by \citet{Garaud2013} and \citet{Drazkowska2014}, but we repeat them here for clarity. Considering first the models without growth by mass-transfer, we see that the mass-distribution is monotonically decreasing for sizes beyond the peak. This small number of particles beyond the peak size already a sign of lucky growth, when compared with models that do not include the full distribution of collision velocities (c.f. \citealt{Windmark2012b,Garaud2013}), which instead show a sharp cut off in the mass-distribution due to a sharp transition from growth to fragmentation. However, because the mean collision velocity continually increases above the bouncing threshold the probability of growth continually decreases while the probability of fragmentation increases, resulting in a monotonic decrease in the number of particles with size.
Conversely, in cases that include mass-transfer, we see that once particles reach a given size (a few cm in this model), the most likely outcome becomes growth again. We will say that particles that have successfully reached this region have broken through the bouncing barrier. Finally, it is worth noting that the maximum size of particles that have broken through will eventually be limited by removal of the largest grains by radial drift \citep{Birnstiel2010,Estrada2016}, which is not included in our local models.}

\reply{Although the mass-transfer models with $\phi=100$ show breakthrough (\autoref{Fig:MTC}), we see that the level of breakthrough is very small in both cases, so much so that no particles can be expected to breakthrough within the lifetime of the disc. Both methods are able to identify that breakthrough is negligible in this case, although it is clear that the level of breakthrough produced in the coarse sub-grid model is entirely dominated  by diffusive growth. In comparison, the fine sub-grid model underestimates the amount of breakthrough.}

\begin{figure}
 \centering
\includegraphics[width=\columnwidth]{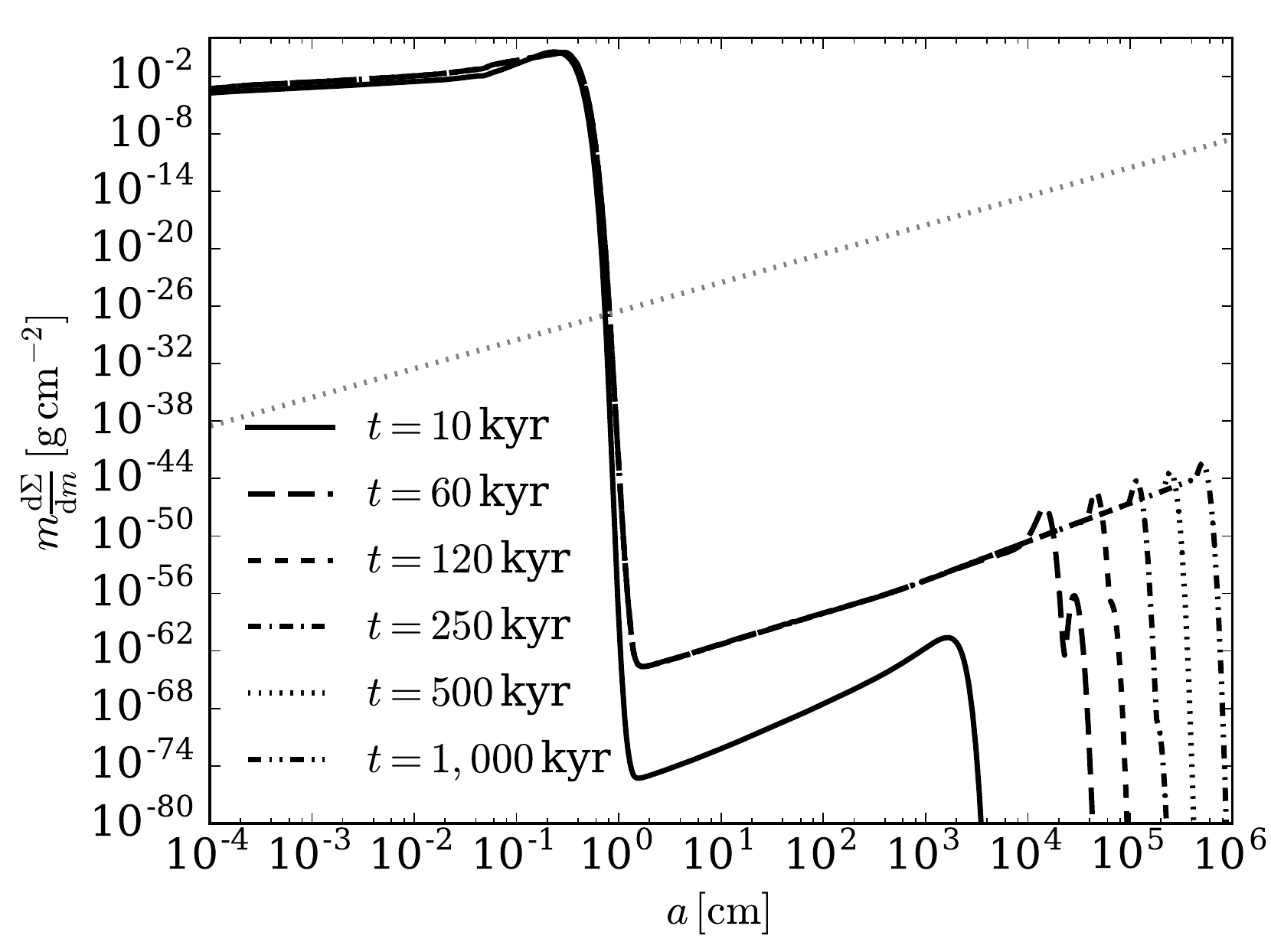} 
\caption{Time evolution of the standard breakthrough model, computed with the fine sub-grid method and $N_{\rm bd}=20$. The grey dashed line denoted 1 particle per unit logarithmic internal in mass and disc radius.}
\label{Fig:phi100_evo}
\end{figure}

\reply{For this choice of parameters we are not able to find convergence with either method, but the fine-grid shows considerably less variation which suggests that it is closer to the true solution. To confirm this, we computed the mean density of particles per logarithmic interval for sizes between 2 and $100\unit{cm}$ and performed Richardson extrapolation on the three highest resolution simulations available for each method (see \autoref{Fig:MTC}) to estimate the level of breakthrough that the two methods are converging to. This produced estimates of $10^{-62}$ for the coarse sub-grid method and $10^{-68}$ for the fine-grid method. Although these estimates are still quite different, given the vast difference between the numerical results, we can have some confidence that the two methods are converging towards the same solution and that the fine sub-grid model does a better job of reproducing it.} In \autoref{Fig:phi100_evo} we show the time evolution of the fiducial model. We see that within about $6\times 10^{4}\unit{yr}$ the bulk of the distribution has reached a steady state, with $n(m)$ in the breakthrough region being constant in time. Beyond the transition size particles grow by sweeping up smaller particles, whereas below the transition collisions result in fragmentation.

The form of the steady-state distribution for the bulk of the mass distribution is due to self-similar growth. \citet{Birnstiel2011} explained the form of the bulk of the distribution as follows: the mass-flux of growing particles is constant with the small particles being replenished by fragmentation. Similar arguments also apply in the power law tail of particles beyond the bouncing barrier. Since in this region the particle density is very low collisions between similarly sized particles can be neglected and growth occurs by sweeping up small particles. The growth rate of the particles is then simply proportional to their area, $\deriv{m}{t} \propto m^{2/3}$, which produces a $n(m) \propto m^{-2/3}$ distribution. The maximum size of particles that have broken through continually increases according to $m(t) \propto t^3$. From our simulations we measure $n(m) \propto m^{-0.65}$, in good agreement. We note that this slope is different to the $n(m) \propto m^{-1.51}$ that \citet{Lee2000} find for the same cross-section and constant collision velocities, but the growth rate is the same. The different slope should be expected since here there is an essentially infinite reservoir of small particles to drive the growth while in \citet{Lee2000} growth of the bulk of the distribution occurs predominantly through collisions with similar sized objects.

\begin{figure}
 \centering
\includegraphics[width=\columnwidth]{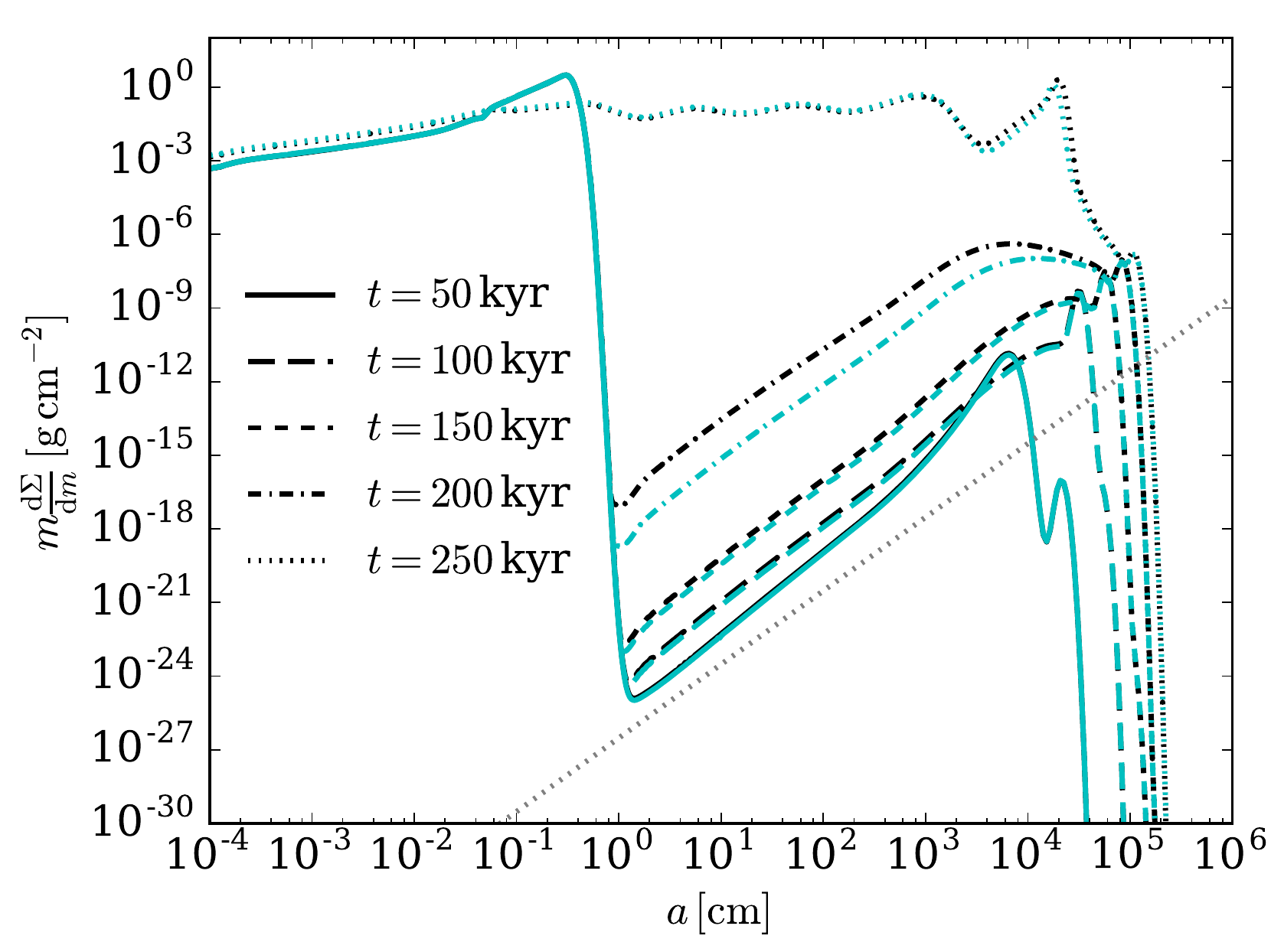} 
\caption{Particle distribution at $1\unit{au}$ for the full mass transfer model with the standard parameters, except with $\phi=50$. The model is computed with  $N_{\rm bd} = 30$ (black) and $N_{\rm bd} = 40$ (blue). Above the bouncing barrier $\deriv{N}{m}$ continues to increase until breakthrough occurs after $2.0 \times 10^{5}\unit{yr}$. 
}
\label{Fig:MT_phi50}
\end{figure}

In contrast at $\phi = 50$ we see a different behaviour. In this case we eventually observe wholesale breakthrough after about $200,000\unit{yr}$, as shown in \autoref{Fig:MT_phi50}. In conjunction with breakthrough, the fragmentation rate increases leading to a larger population of small grains. We note that the structure seen in $m\deriv{\Sigma}{m}$ at the largest sizes remains from the initial transient that occurs as the first particles break through the bouncing barrier (seen at a few $10^4\unit{cm}$ after $5 \times 10^4\unit{yr}$ in \autoref{Fig:MT_phi50}). While the shape of this feature is not dependent on resolution, we do not ascribe any importance to it as it may depend on the initial conditions. 

Comparing the two resolutions shown in \autoref{Fig:MT_phi50}, we see excellent agreement between the solutions at $N_{\rm bd} = 30$ and $N_{\rm bd} = 40$ at $5 \times 10^{4}\unit{yr}$. Similarly, at $5 \times 10^{4}\unit{yr}$ even lower resolutions (not shown) are also in excellent agreement. However, when the two models are close to breakthrough the differences are larger, with the lower resolutions breaking through earlier. After breakthrough once again the two solutions are in excellent agreement. \reply{Given the good agreement before and after breakthrough, this suggests that diffusive growth should not affect the long term behaviour when this type of wholesale breakthrough occurs.} 

\begin{figure}
 \centering
\includegraphics[width=\columnwidth]{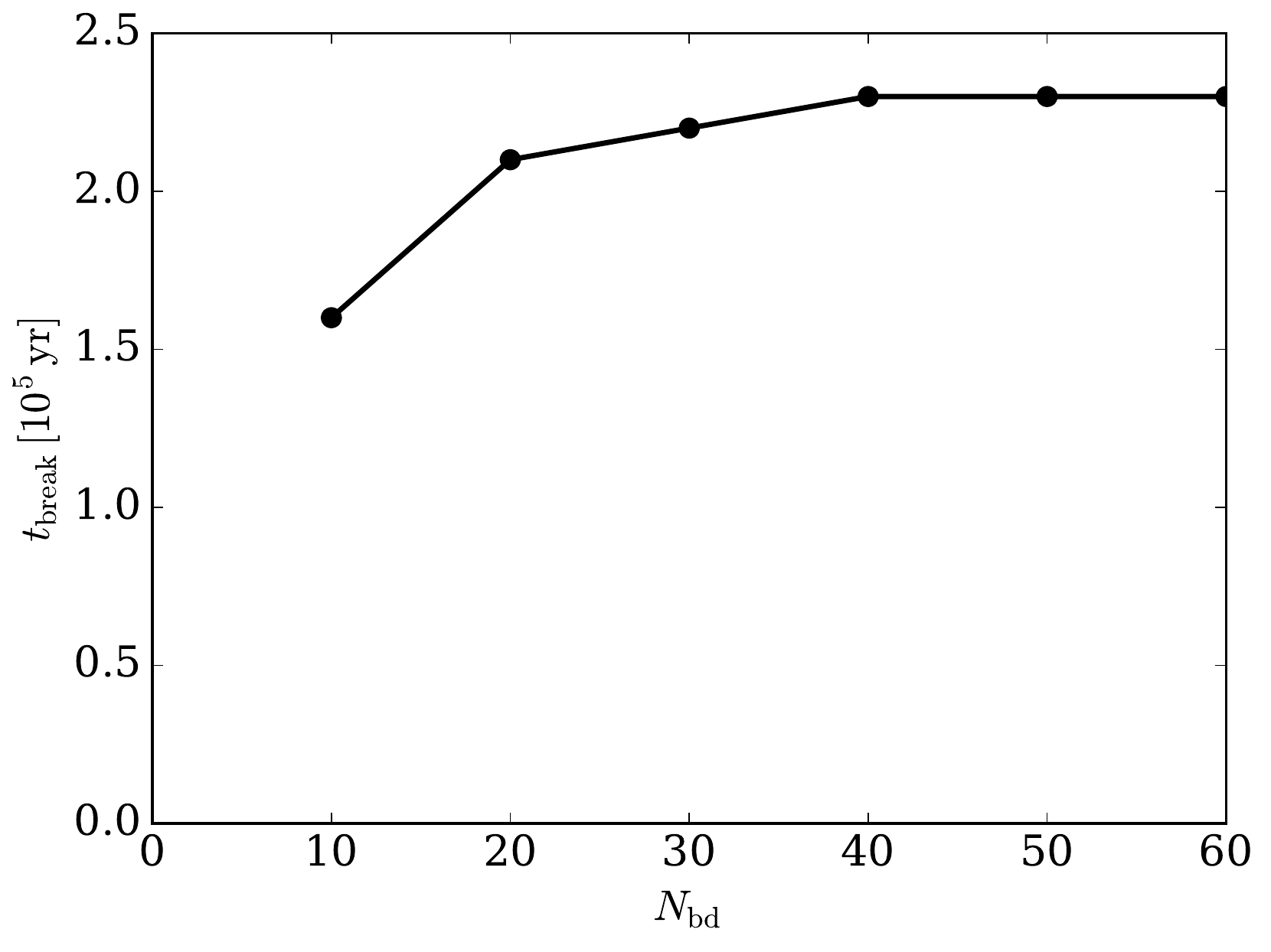} 
\caption{Breakthrough time for the same model as \autoref{Fig:MT_phi50} as a function of resolution.}
\label{Fig:t_break_res}
\end{figure}

Similarly to \citet{Drazkowska2014}, to quantify the resolution dependence and convergence in breakthrough models we use the time at which breakthrough occurs, $t_{\rm break}$. We define this to be the time at which more than half the mass is in particles of size $s > 10\unit{cm}$. \autoref{Fig:t_break_res} demonstrates the resolution dependence of $t_{\rm break}$. While there is some resolution dependence in $t_{\rm break}$, we see that at modest resolution ($N_{\rm bd} = 20$) the error is already less than ten percent. We have not computed $t_{\rm break}$ using the coarse method due to the long integration time and additional computational cost of the coarse method\reply{; however, Figures 2 \& 3 in \citet{Drazkowska2014} show a similar error for $N_{\rm bd} = 10$ (approximately a  factor of 2).}

\begin{figure}
 \centering
\includegraphics[width=\columnwidth]{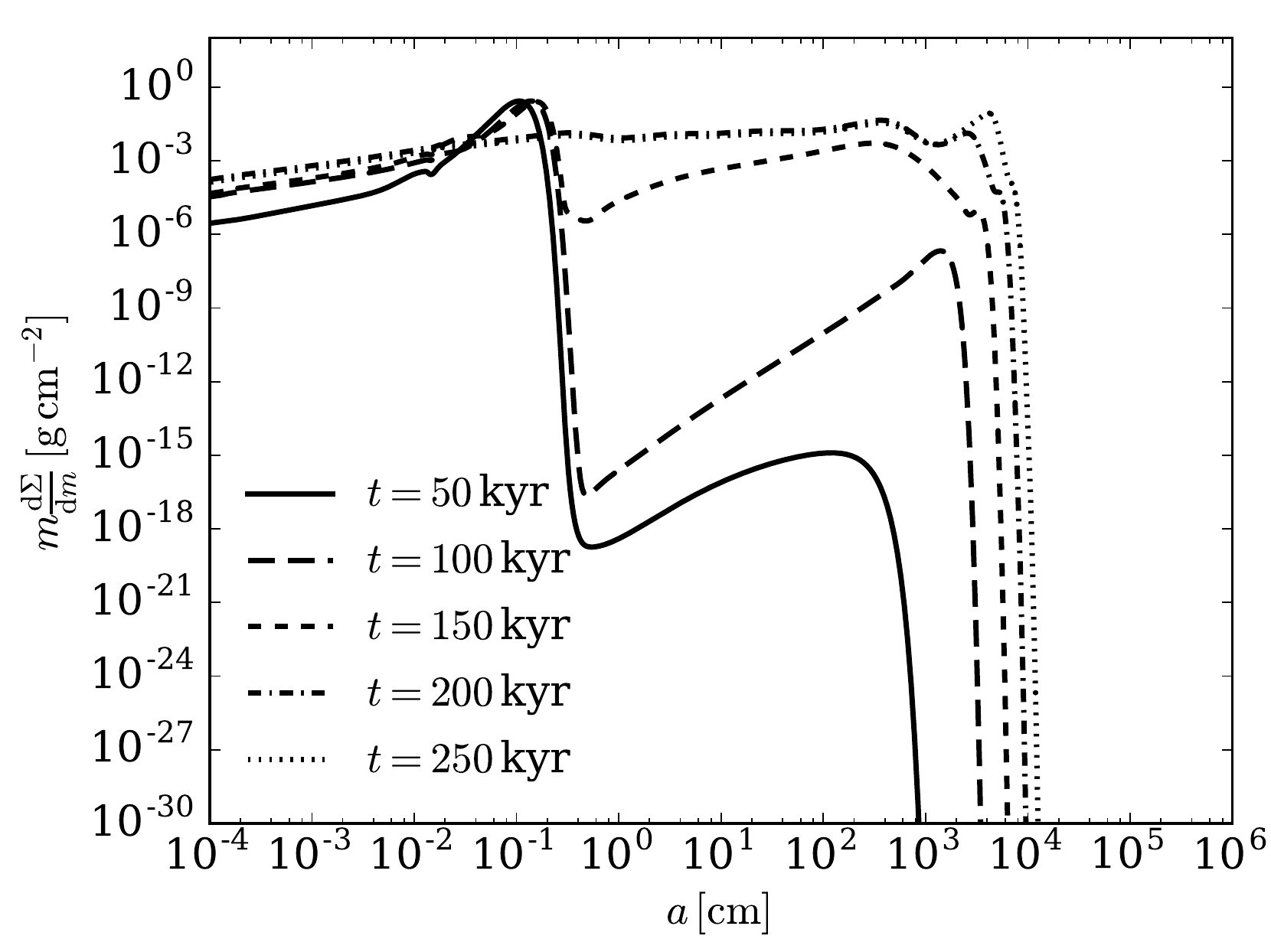} 
\caption{Breakthrough at $R = 3\unit{au}$ with  $\phi = 30$. The solution shown is computed for $N_{\rm bd} = 20$.}
\label{Fig:MT_phi30}
\end{figure}

To demonstrate why reducing $\phi$ leads to breakthrough we show another model for $\phi = 30$ and $R = 3\unit{au}$. In this model (\autoref{Fig:MT_phi30}) it is easy to see that the slow increase in the number density in the tail occurs along with a gradual shift of the peak of the distribution (near $1\unit{mm}$) to larger sizes and an increase in number of particles at small sizes. Thus lucky growth has two effects: 1) a few particles are able to grow large enough to be able to sweep up small particles and grow via mass transfer, essentially by luckily avoiding fragmenting collisions and 2) as the peak of the distributions moves to larger sizes it is able to slowly grow via an increasingly small contribution of low velocity collisions. As 2) occurs, the mean collision velocity increases, increasing the population of small grains produced by fragmentation. This additionally increases the rate of growth by mass transfer, eventually leading to rapid breakthrough.

\begin{figure*}
\centering
\includegraphics[width=\textwidth]{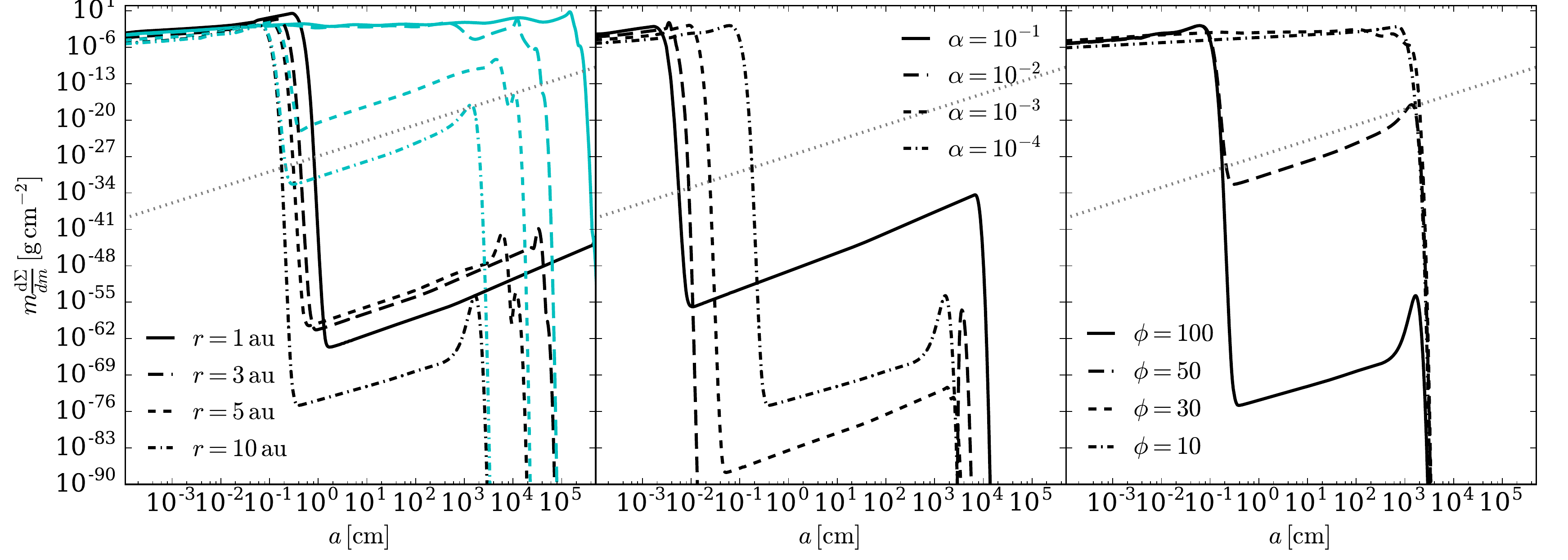} 
\caption{Mass distribution at $10^6\unit{yr}$ for different models. All models are computed at a resolution of $N_{\rm bd} = 20$ and use the canonical parameters, \reply{but at $R=10\unit{au}$. In each panel one of either $R$, $\alpha$, or $\phi$ are varied. The grey dotted line denotes 1 particle per logarithmic interval in mass and radius at $10\unit{au}$. In the left panel, both $\phi = 100$ (black) and $\phi = 50$ (cyan) are shown.} The critical mass-ratio $\phi$ is much more important for determining whether breakthrough occurs than the other parameters.}
\label{Fig:model_grid}
\end{figure*}

We now turn our attention to how the model parameters affect the distribution of particle masses by varying $R$, $\alpha$ and $\phi$. In \autoref{Fig:model_grid} we show the results of models at $N_{\rm bd}=20$. We see that while varying $R$ and $\alpha$ affects the bulk of the distribution and the level of breakthrough in the tail of the distribution, only $\phi$ affects whether breakthrough occurs. The effects of varying $R$, and $\alpha$ are consistent with results found in previous works \citep[e.g][]{Brauer2008,Windmark2012b,Garaud2013}: increasing $R$ leads to particles reaching the bouncing barrier at smaller sizes since the increase in $\St$ (which increases the collision velocity) outweighs the decrease in $c_s$ and $\Omega$ (which decrease the collision velocity), while increasing $\alpha$ leads to higher collision velocities and hence faster growth but smaller sizes at the bouncing barrier.

\reply{The higher level of breakthrough at $\alpha = 10^{-1}$ shown in \autoref{Fig:model_grid} shows a change in regime. For such strong turbulence the fragmentation barrier appears at the boundary between the turbulence dominated and Brownian motion dominated regimes, where there is a large jump in the collision velocity \citep[e.g. Figure 3 of][]{Ormel2007}. The reason breakthrough happens more readily in this case is due to the shape of the density distribution. Since there is no bouncing regime, there is no region of slower growth that causes  a bump in the density distribution. This means that growth by mass-transfer is more efficient relative to fragmentation, making breakthrough easier. In this high $\alpha$ regime the size at the fragmentation boundary becomes independent of $\alpha$, instead being controlled by the Reynolds number, which we have taken to be $10^8$.}

\reply{In exploring a full grid of the parameter space shown in \autoref{Fig:model_grid} using the fine sub-grid model, we found that at $\phi = 100$ breakthrough is always restricted to the tails of the distribution, and always at levels that are low enough to be negligible (i.e. the expected number of particles breaking through within $10^6\unit{yr}$ is less than 1). Conversely, for $\phi \lesssim 50$ we found that breakthrough always eventually occurs in the bulk of the distribution (although this can take considerably longer than $10^6\unit{yr}$ at large distances and low $\alpha$).  This gives three types of behaviour for this model. The extremes are that without mass-transfer no breakthrough occurs, and for efficient mass-transfer eventually the whole distribution overcomes the bouncing barrier. In the middle region  a few particles in the tail breakthrough; however, the range of $\phi$ in which this behaviour results in a significant number of particles growing beyond the bouncing barrier is small (${100 \lesssim \phi \lesssim 50}$).}

While \emph{quantitative} agreement with previous studies is not possible because they were either affected by excessive diffusive growth or used a different kernel, we can compare qualitatively our types of behaviour to previous studies on breakthrough. \reply{First, we note that both methods produce the same phenomenological behaviour in all cases, i.e they agree whether breakthrough occurs in the bulk of the distribution, the tails, or not at all.} We find excellent phenomenological agreement with \citet{Garaud2013} who use the same kernel as us and the coarse sub-grid approach at a lower resolution. For example, they also find wholesale breakthrough at $\phi = 50$, while at $\phi=100$ they find that breakthrough only occurs in the tails, albeit at a much higher level than we do. Comparison with \citet{Windmark2012b} is more difficult as they only present a single model at $50,000\unit{yr}$ and with a different velocity law. In our closest model to theirs, i.e. with $\phi=50$, we find wholesale breakthrough after $200,000$ to $250,000\unit{yr}$, and also find  higher levels of breakthrough at $50,000\unit{yr}$. Since our method produces lower diffusion than the coarse sub-grid method (which they use) the higher level of breakthrough likely reflects that breakthrough is easier when accounting for the full distribution than using a Maxwellian approximation, as found by \citet{Garaud2013}. Similarly, we agree with \citet{Drazkowska2014} in the resolution needed to accurately resolve the breakthrough time. The relatively good qualitative agreement that we find with these previous works shows that their conclusions are robust despite the numerical problems they encountered.

\subsection{Physically motivated model}

\begin{figure}
\centering
\includegraphics[width=\columnwidth]{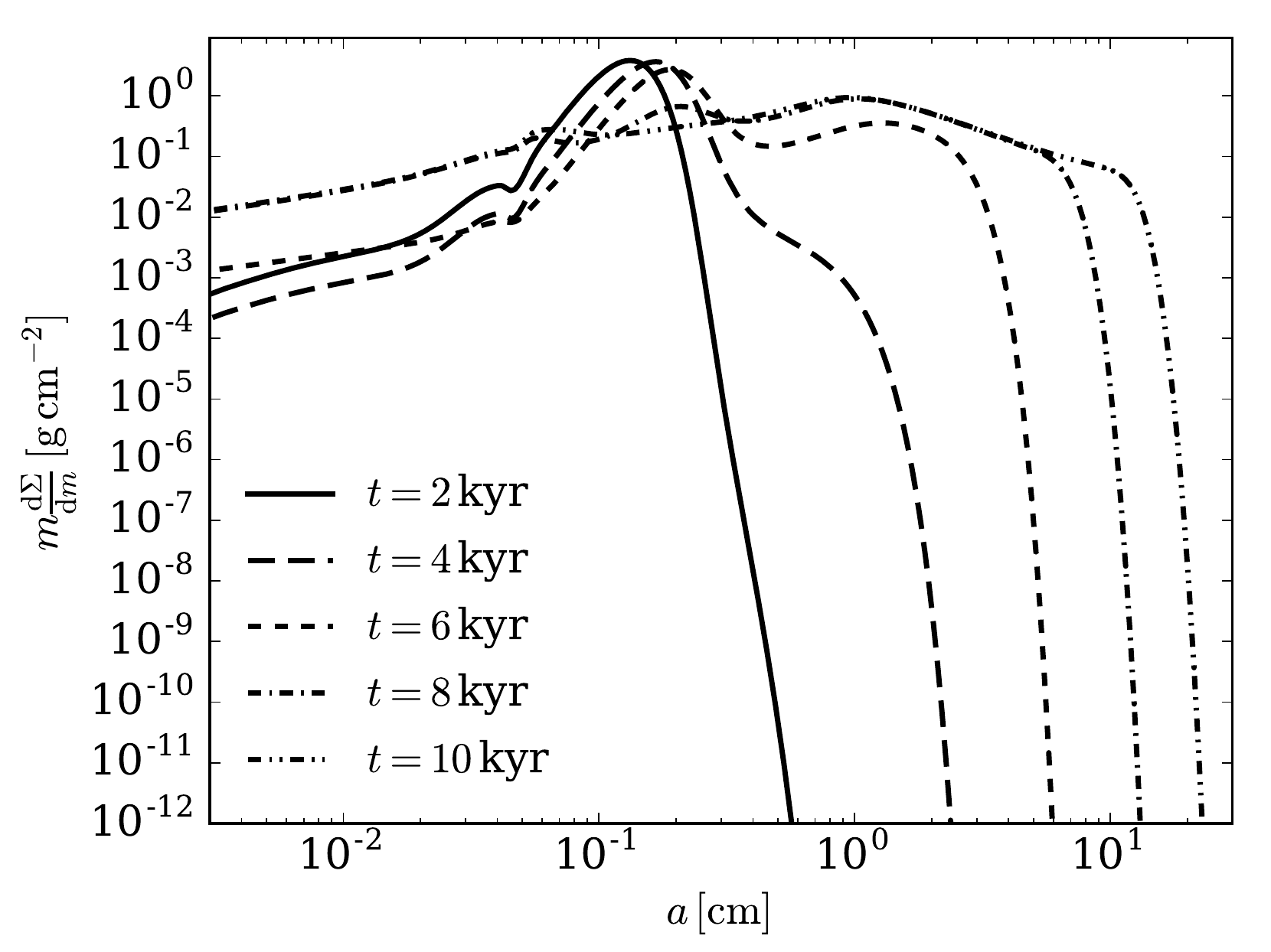} 
\caption{Evolution of the mass distribution for the physically motivated fragmentation model, with $\epsilon^{\rm m} = 0.1$, at $1\unit{au}$, computed using the fine sub-grid model and $N_{\rm bd}=20$.}
\label{Fig:phys_evo}
\end{figure}

\begin{figure}
\centering
\includegraphics[width=\columnwidth]{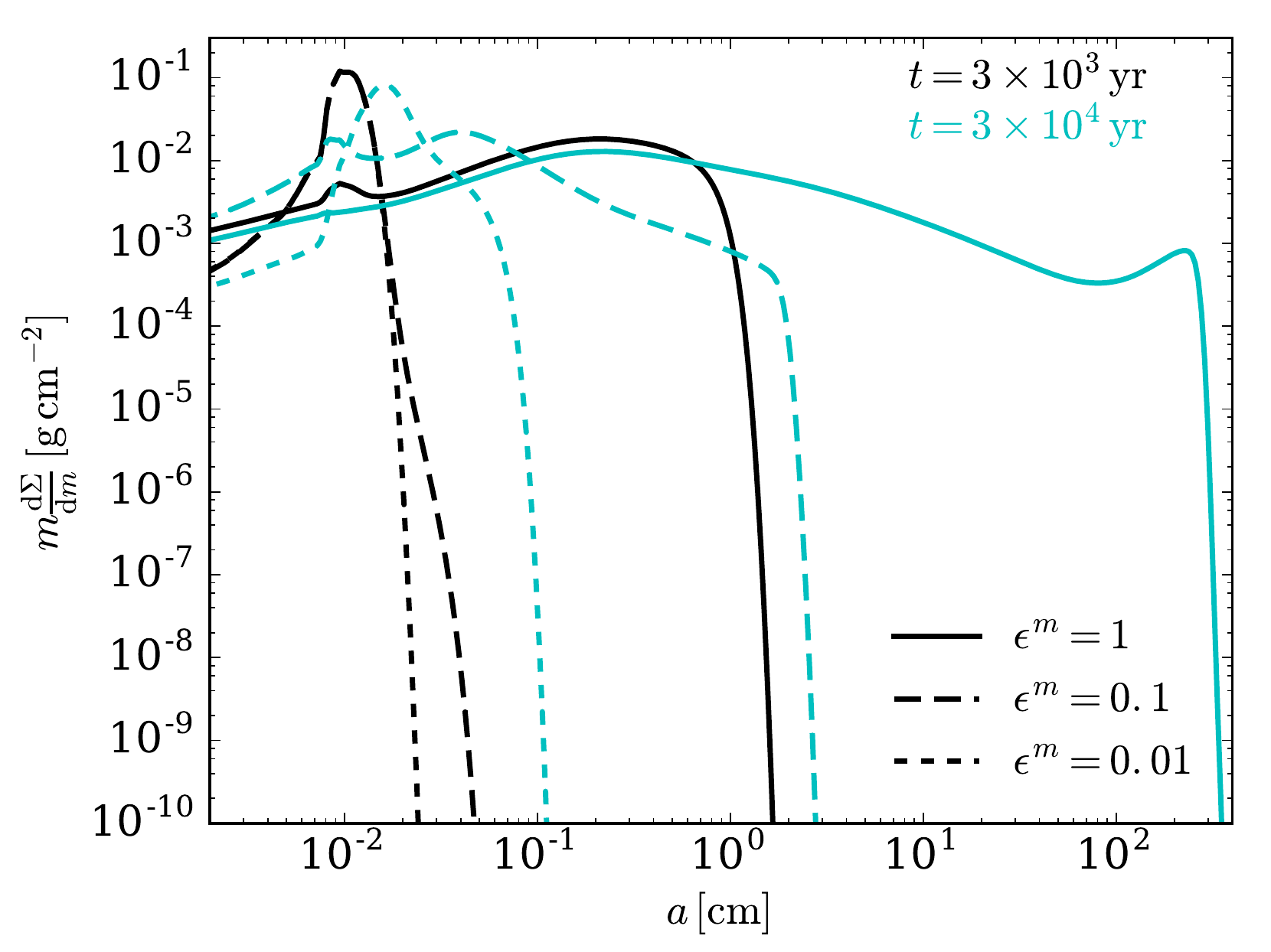} 
\caption{Mass distribution for different mass transfer efficiencies at two different times, for a model with $\alpha = 10^{-3}$ and $R=5\unit{au}$. Models were computed with the fine sub-grid model and $N_{\rm bd} = 20$. At $3\times 10^4\unit{yr}$ the model with $\epsilon^m=0.01$ is just beginning to break through the bouncing barrier.}
\label{Fig:eps_compare}
\end{figure}

\reply{We now consider how the physically motivated fragmentation and mass-transfer model affects lucky growth and breakthrough. In \autoref{Sec:Tests} we already saw that the total fragmentation rate can be much lower than in the simple model, and thus we expect breakthrough to occur more readily. In \autoref{Fig:phys_evo} we show the evolution of the default model, but using the physically motivated fragmentation and mass-transfer model, with $\epsilon^{\rm m} = 0.1$. The evolution again follows the pattern of rapid growth up to mm sizes, which stalls and is then followed by breakthrough.}

\reply{As in the simple model with low values of the critical mass-transfer ratio, breakthrough is driven by a combination of the slow increase in the peak grain size and mass-transfer in the tail. This is illustrated in \autoref{Fig:eps_compare}, where $\epsilon^{\rm m}$ has been varied. At $3 \times 10^3\unit{yr}$ the models with $\epsilon^{\rm m} = 10^{-2}$ and 0.1 have the same density at the peak size ($\sim 10^{-2}\unit{cm}$), but the number of particles in the  tails differ. Once the number of particles that breakthrough becomes significant the evolution at the peak begins to differ as well. This demonstrates that mass-transfer is not affecting the evolution of the peak size prior to wholesale breakthrough. However, the mass-transfer efficiency does affect the time taken to breakthrough, though fairly weakly with tests over a range of conditions showing the breakthrough time scales as $\sim(\epsilon^{\rm m})^{1/2}$. Conversely, the growth-rate once breakthrough has occurred is more sensitive to the mass-transfer efficiency, as can be seen by the maximum size of particles in \autoref{Fig:eps_compare}.}

\begin{figure}
\centering
\includegraphics[width=\columnwidth]{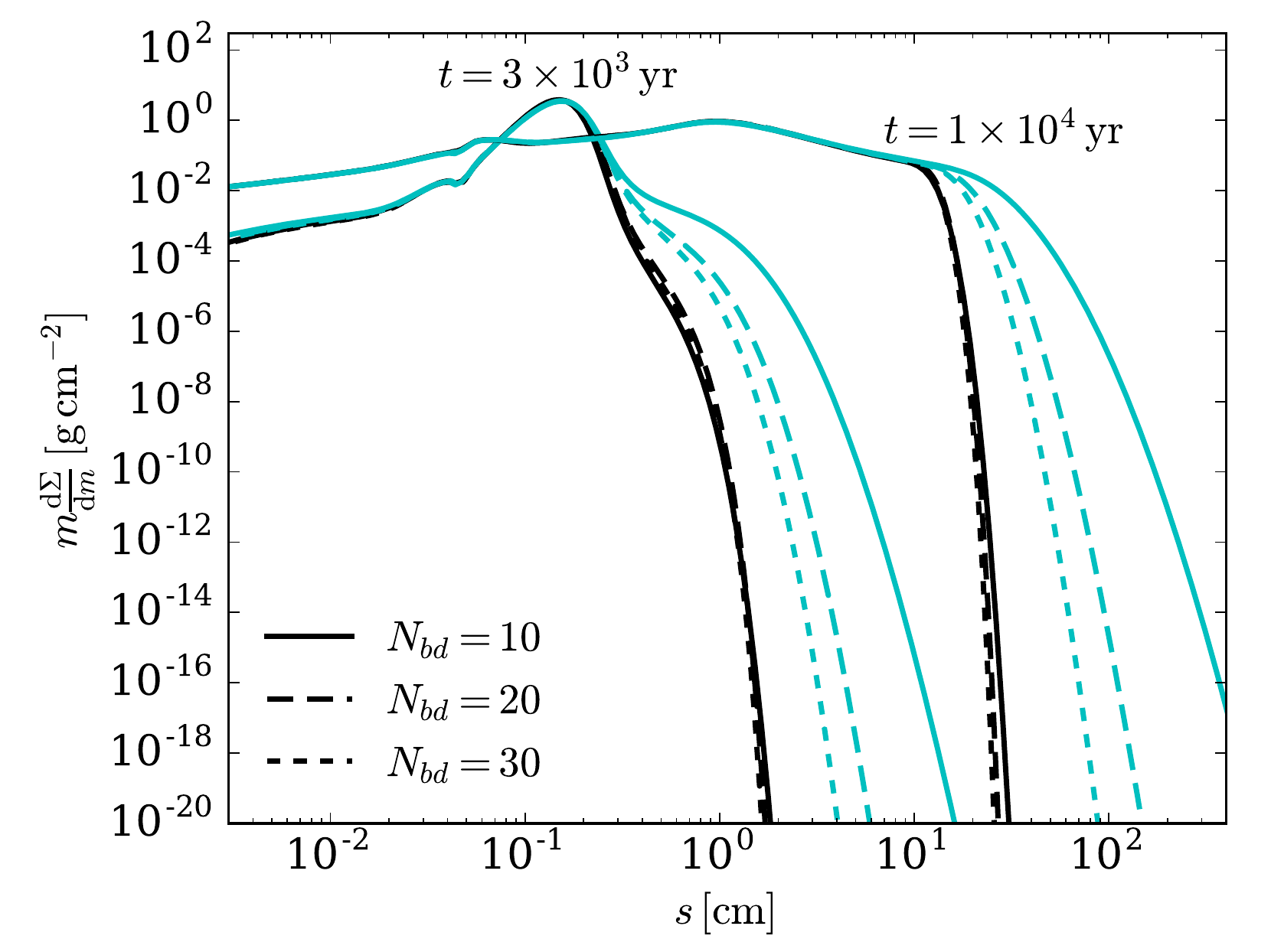} 
\caption{Convergence of the mass distribution for the physically motivated fragmentation model (with $\epsilon^m = 0.1$) using both the fine (black) and coarse (cyan) sub-grid models at two times.}
\label{Fig:phys_compare}
\end{figure}

\reply{Similarly to the simple fragmentation model, the fine sub-grid model is more effective at resolving the density structure, as shown in \autoref{Fig:phys_compare}. Both methods are again effective at capturing the phenomenological behaviour, but as before the coarse sub-grid model shows artificially enhanced growth. This results in slightly earlier breakthrough, which leads to the turnover in the density appearing at larger sizes. This difference persists as the evolution continues. At higher resolution the differences are smaller with the coarse sub-grid model converging towards the fine sub-grid model.}

\reply{We have also explored whether this model ever gives rise to the type of growth seen in the simple model with large $\phi$: the bulk of the mass remaining below the bouncing barrier, but with a small number of particles able to breakthrough and grow to large sizes. However, we find that reducing the mass-transfer efficiency, $\epsilon^{m}$, does not produce this behaviour, instead it just delays the wholesale breakthrough and slows subsequent growth. We also experimented with re-introducing the critical mass ratio, $\phi$, such that only collisions between particles of mass ratio greater than $\phi$ are considered for growth by mass transfer, again finding that this increases the breakthrough time without changing the phenomenological behaviour.}

\section{Discussion}
\label{Sec:Discuss}

The tendency of grains to bounce off of each other even at relatively low velocity (a few $\mathrm{cm\,s}^{-1}$ when compact, \citealt{Seizinger2013a}) presents a barrier to growth that can trap grains at millimetre to centimetre sizes. The extra stickiness of icy grains in the outer disc means that they may remain porous and  avoid the bouncing barrier \citep{Okuzumi2012,Krijt2015}, thus bouncing is likely most important in the inner disc. \reply{However, it has been suggested that this barrier may be overcome through a combination of rare low velocity collisions between particles that result in growth, and mass-transfer, a process in which one of the colliding particles fragments and is partially accreted by the other \citep{Windmark2012a,Windmark2012b,Garaud2013,Estrada2016}.} 

\reply{The efficiency of these processes in overcoming the bouncing barrier has, however, remained uncertain. This is in part due to the previous studies lacking the necessary resolution to accurately determine the outcome, further exacerbated by overly simple growth and fragmentation models usually employed. The simple models employed assume that collisions above a fixed threshold velocity resulted in fragmentation, unless the mass-ratio between the colliding particles was above some critical threshold, $\phi$, in which case the result was instead growth by mass-transfer. To resolve the numerical issues we have conducted a detailed study of breakthrough using a pair of numerical approaches, including a standard Smoluchowski equation based approach \citep{Brauer2008,Garaud2013} and a rapidly converging variant based upon \citet{Lee2000} to separate physical behaviour from numerical artefacts.}

\reply{Here, we investigated the overly optimistic case in which mass-transfer is assumed to be perfectly efficient, i.e. high velocity collisions between pairs of particles above the threshold result in sticking, whereas those below it result in fragmentation. These two numerical methods show that despite difficulties achieving numerically converged results, the phenomenological behaviour seen in previous studies is robust \citep{Windmark2012b,Garaud2013,Drazkowska2014}. These results can be summarized into three behaviours. Without mass-transfer a few lucky particles may grow beyond the bouncing barrier, but never to large sizes. With mass-transfer and $\phi \gtrsim 100$, fragmentation is so efficient that growth far beyond the bouncing barrier is limited to a few lucky particles that managed to grow without fragmenting. Using high resolution models we determined that the number of particles that manage to proceed grow this way is negligibly small, and thus growth essentially stops at the bouncing barrier in this region too. For $\phi \lesssim 50$, the phenomenological behaviour was entirely different: growth beyond the bouncing barrier proceeds initially with a small number of particles, but this continually increases until the whole distribution of particles breakthrough.}

\reply{The above results would appear to suggest that growth beyond the bouncing barrier is not important in general because the mass-transfer efficiency is typically much smaller than 100 per cent. Numerical and experimental results suggest that the efficiency of mass transfer is typically only around 10 per cent \citep{Beitz2011,Windmark2012a,Seizinger2013a}. Whether a lower mass-transfer efficiency would require a lower $\phi$ for breakthrough in the bulk of the distribution is not immediately clear, because this form of breakthrough is associated with the slow increase in the average size of the particles, which can occur due to lucky growth of particles at the bouncing barrier. However, $\phi$ clearly affects the rate at which an average size particle grows (compare \autoref{Fig:MT_phi50} with \autoref{Fig:MT_phi30}, which breaks through earlier despite the lower collision rate at $3\unit{au}$), and the time taken for breakthrough (which can be seen from \autoref{Fig:model_grid}, where $\phi=30$ has broken through at $10^6\unit{yr}$ but $\phi=50$ has not). This suggests that a smaller critical $\phi$ may be needed with a lower mass-transfer efficiency.}

\reply{However, \citet{Estrada2016} found significant levels of breakthrough (albeit never in the bulk of the mass distribution) in global models that were based on a physical prescription for mass transfer and fragmentation following \citet{Windmark2012a}, that also included erosion and  are based upon experimental data \citep{Guttler2010}. This result is surprising in the context of the results above due to the lower mass-transfer efficiency. To explore this, we have also investigated breakthrough in a model that includes a more physically motivated fragmentation model, which takes into account the dependence of the fragmentation threshold on the mass ratio of the colliding particles (see \autoref{Sec:Kernel}). The essence of this is that when the mass ratio is large, collisions above the standard fragmentation threshold may fragment the smaller particle, without having the energy to fragment the larger one, resulting in a higher fragmentation threshold for the larger particle. Collisions between the two thresholds were assumed to result in mass-transfer, with some efficiency, $\epsilon^{m} \le 1$. This greatly reduces the fragmentation rate between pairs of particles with large mass ratios.}

\reply{The lower fragmentation rate of particles near the bouncing threshold in the physically motivated model means that breakthrough can occur much more readily. Our models showed that even at low mass-transfer efficiencies breakthrough can eventually occur in the bulk of the distribution. However, we note that for low mass-transfer efficiencies, breakthrough will be prevented by radial drift removing the largest particles before they grow large enough because the radial drift time-scale is typically shorter than the growth time-scale for Stokes numbers close to unity \citep{Birnstiel2012}, except perhaps in the case of porous growth (\citealt{Krijt2016}). This model shows the prospects for breakthrough remain more optimistic than the simple models would suggest, and  that differences between such simple models and the global models of \citet{Estrada2016}  are attributable to a different treatment of fragmentation.} 

\reply{The physically motivated models presented here, along \citet{Estrada2016} show that the bouncing barrier is likely `soft', i.e. given enough time particles are able to grow beyond it until limited by some other process. In smooth discs this is likely the removal of large grains by radial drift. However, in particle traps where radial drift is prevented growth to large sizes may be possible. Finally erosion, which we have also not included, also likely plays an important role for determining how large the grains are able to grow.}

\reply{Recent experimental work supports the idea of growth beyond the bouncing barrier. \citet{Kruss2017} found that even though large particles do not grow through low velocity collisions because the aggregate is too weak and breaks apart again. They also showed that large particles at the bouncing barrier are able to still grow by sweeping up small particles. Since infrequent high velocity collisions are able to maintain a reservoir of small particles \citep[see, e.g.][]{Garaud2013}, particles at the bouncing barrier will be able to continue to grow. Thus the prospects for growth beyond the bouncing barrier to larger sizes are good.}

\section{Conclusions}
\label{Sec:Conclude}

\reply{Using a high precision Smoluchowski equation approach to coagulation and fragmentation (based on \citealt{Lee2000}) we have revisited the role of bouncing, fragmentation and mass transfer in the evolution of compact grains that are close to the bouncing barrier (mm or cm sizes). Taking into account the full distribution of collision velocities we have investigated the role of lucky growth in the evolution, i.e. whether particles can grow beyond the bouncing barrier through infrequent collisions at low velocity despite the high mean collision velocity.}

\reply{Our models confirm previous work based on simple models for mass-transfer and fragmentation, which found that growth by mass transfer, in which one particle is fragmented and partially accreted by the other, can help particles grow beyond the bouncing barrier \citep[e.g.][]{Garaud2013,Windmark2012b,Drazkowska2014}. Using a numerical approach with low diffusion, we were able to demonstrate that typically the number of particles breaking through the bouncing barrier is negligibly low unless the critical mass ratio, $\phi$, above which high velocity collisions result in growth rather than fragmentation is low ($\phi \lesssim 50$). Given that we assumed an overly efficient model for mass-transfer, this appeared to suggest that growth beyond the bouncing barrier is unlikely.}

\reply{However, in a second model that includes the fact that the fragmentation threshold increases with mass-ratio, we found that growth beyond the bouncing barrier can occur much more readily. The differences arise because the slow growth beyond the bouncing barrier is in competition with fragmentation. Since fragmentation requires higher velocities at higher mass-ratio, this model produces a lower fragmentation rate and thus growth is much easier.}

\section{Acknowledgements}
We thank Pascale Garaud for several useful discussions and making her code available. We thank the reviewer, Paul Estrada, for his detailed comments which led to an improved manuscript. This work has been supported by the DISCSIM project, grant agreement 341137 funded by the European Research Council under ERC-2013-ADG and the Hong Kong RGC grant HKU 7030/11P. FM acknowledges support from the Leverhulme Trust Early Career Fellowship and the Isaac Newton Trust. This work used the DIRAC Shared Memory Processing system at the University of Cambridge, operated by the COSMOS Project at the Department of Applied Mathematics and Theoretical Physics on behalf of the STFC DiRAC HPC Facility (www.dirac.ac.uk). This equipment was funded by BIS National E-infrastructure capital grant ST/J005673/1, STFC capital grant ST/H008586/1, and STFC DiRAC Operations grant ST/K00333X/1. DiRAC is part of the National E-Infrastructure.

\bibliography{breakthrough}
\bibliographystyle{mnras_edit}

\bsp

\appendix

\label{lastpage}
\end{document}